\documentclass[a4paper,11pt]{article}
\usepackage{jheppub}

\usepackage{xcolor}
\usepackage{orcidlink}
\usepackage{comment,slashed}

\newcommand{\MeV}{\,\mathrm{MeV}}

\newcommand{\erg}{\,\mathrm{erg}}

\newcommand{\mchi}{m_\chi}

\newcommand{\Rccsn}{R_\mathrm{CCSN}}
\newcommand{\Rsfr}{R_*}

\definecolor{darkred}{rgb}{0.6,0.0,0.}

\title{\boldmath Energy-dependent Boosted Dark Matter from Diffuse Supernova Neutrino Background}

\author[a]{Anirban Das\,\orcidlink{0000-0002-7880-9454},}
\affiliation[a]{Center for Theoretical Physics, Department of Physics \& Astronomy, Seoul National University, Seoul 08826, South Korea}
\emailAdd{anirbandas@snu.ac.kr}

\author[b]{Tim Herbermann,}
\emailAdd{tim.herbermann@mpi-hd.mpg.de}
\affiliation[b]{Max-Planck-Institut für Kernphysik,
Saupfercheckweg 1, 69117 Heidelberg, Germany}

\author[b]{Manibrata Sen\,\orcidlink{0000-0001-7948-4332},}
\emailAdd{manibrata.sen@mpi-hd.mpg.de}
\author[c,d,e,f]{and Volodymyr Takhistov}
\affiliation[c]{International Center for Quantum-field Measurement Systems for Studies of the Universe
and Particles (QUP), KEK, 1-1 Oho, Tsukuba, Ibaraki 305-0801, Japan
}
\affiliation[d]{Theory Center, Institute of Particle and Nuclear Studies, High Energy Accelerator Research Organization (KEK), Tsukuba 305-0801, Japan}
\affiliation[e]{Graduate University for Advanced Studies (SOKENDAI),
1-1 Oho, Tsukuba, Ibaraki 305-0801, Japan
}
\affiliation[f]{Kavli Institute for the Physics and Mathematics of the Universe (WPI), The University of Tokyo Institutes for Advanced Study, The University of Tokyo, Kashiwa, Chiba 277-8583, Japan}
\emailAdd{vtakhist@post.kek.jp}

\abstract{
Diffuse neutrinos from past supernovae in the Universe present us with a unique opportunity to test dark matter (DM) interactions. These neutrinos can scatter and boost the DM particles in the Milky Way halo to relativistic energies allowing us to detect them in terrestrial laboratories. 
Focusing on generic models of DM-neutrino and electron interactions, mediated by a vector or a scalar boson, we implement energy-dependent scattering cross-sections and perform detailed numerical analysis of DM attenuation due to electron scattering in-medium while propagating towards terrestrial experiments. We set new limits on DM-neutrino and electron interactions for DM with masses in the range $\sim (0.1, 10^4)$~MeV, using recent data from XENONnT, LUX-ZEPLIN, and PandaX-4T direct detection experiments. We demonstrate that consideration of energy-dependent cross-sections for DM interactions can significantly affect constraints previously derived under the assumption of constant cross-sections, modifying them by multiple orders of magnitude.}

\begin{document}

\preprint{KEK-QUP-2024-0007, KEK-TH-2610, KEK-Cosmo-0342, IPMU24-0009}
\maketitle
\flushbottom

%%%%%%%%%%
%%%%%%%%%%%
\section{Introduction}
\label{sec:intro}
%%%%%%%%%%%
%%%%%%%%%
Dark matter (DM) constitutes about $\sim 85\%$ of all matter in the Universe. However, despite decades of searches, its nature remains mysterious (see e.g.~\cite{Gelmini:2015zpa} for review).
Traditional direct DM detection experiments, which constitute a cornerstone of DM exploration, have aimed at searching for energy deposits from Galactic halo DM interactions with nucleons (see e.g.~\cite{Akerib:2022ort} for an overview). Recent results from experiments like XENON~\cite{XENON:2023cxc}, LUX-Zepelin (LZ)~\cite{LZ:2022lsv,LZ:2022ysc} and PandaX~\cite{PandaX:2022ood} continue to further push this frontier.
However, scattering with electrons can provide unique opportunities for DM exploration, particularly for sub-GeV low-mass DM which would be kinematically challenging to detect with nuclear scattering. DM-electron scattering has already been adopted as a target for a multitude of experiments, including XENONnT~\cite{XENON:2022ltv},
DAMIC~\cite{DAMIC-M:2023gxo}, SENSEI~\cite{SENSEI:2023zdf} and SuperCDMS~\cite{SuperCDMS:2020ymb} and is expected to become a central avenue for future direct DM detection studies~\cite{Essig:2022dfa}.

Neutrinos are ubiquitously abundant in the Universe, originating from diverse sources such as primordial plasma and astrophysical transients. This provides us with unique opportunities to explore possible neutrino-DM interactions. Scattering of energetic neutrinos with slow-moving cold DM particles within the Milky Way halo can upscatter the latter to relativistic energies. Such boosted DM (BDM) would manifest in distinct experimental signatures and offer new means of detection. One motivated class of DM models to explore within this context is leptophilic DM\,\cite{Bernabei_2008,falkowski2009dark,Fox_2009,Chang_2014,Lindner_2010,macias2015effective,Blennow_2019,FileviezPerez:2019cyn}, where the DM interacts exclusively with neutrinos as well as charged leptons of the Standard Model (SM).
Furthermore, these interactions facilitate DM scattering off free leptons encountered on their journey towards Earth.
We note that analogous boosting of DM can also be achieved in the context of hadrophilic DM models due to energetic hadronic interactions of DM with cosmic rays, which has been extensively explored~\cite{DEramo:2010keq,Agashe:2014yua,Berger:2014sqa,Ema:2018bih,Bringmann:2018cvk,Cappiello:2018hsu,Yin:2018yjn,Kannike:2020agf,Fornal_2020,Alhazmi:2020fju,Das:2021lcr,Jho:2021rmn,Das:2022xsz,Berger:2022cab,Bardhan:2022bdg,Maity:2022exk,DeRomeri:2023ytt, Xia:2024ryt,Guha:2024mjr}. Such DM interactions enable probing lighter sub-GeV DM masses with nucleon scattering by overcoming kinematic thresholds which impede cold non-relativistic DM searches in conventional noble element detectors.

In this paper, we revisit the boosting of light DM by neutrinos contributing to the diffuse supernova neutrino background (DSNB) from historic supernovae events~\cite{Das:2021lcr} and expand the analysis in many aspects. An explosive core-collapse supernova (CCSN) loses a major fraction of its energy through neutrino emission. 
This was confirmed by detection of neutrinos~\cite{Kamiokande-II:1987idp,Bionta:1987qt}
from SN1987A that exploded in the Large Magellanic Cloud, which also spearheaded the era of multimessenger astronomy.
Such neutrinos follow an approximate thermal distribution with energies $E\sim \mathcal{O}(10)\MeV$. Accumulation of neutrinos from all the previous CCSNe in the history of our Universe contributes to the persistent flux of DSNB, also known as supernova relic neutrino background (see e.g.~\cite{Beacom:2010kk, Ando:2023fcc} for review). Detection of the DSNB is a prime target of current as well as forthcoming neutrino experiments, including Super-Kamiokande (SK)~\cite{Zhang:2013tua}, Hyper-Kamiokande (HK)~\cite{Hyper-Kamiokande:2018ofw}, Jiangmen Underground Neutrino Observatory (JUNO)~\cite{JUNO:2015zny}, and Deep Underground Neutrino Experiment (DUNE)~\cite{DUNE:2020ypp}. At present, the most stringent upper limit on the DSNB flux, set by SK, remains a few times higher than typical theoretical predictions~\cite{Super-Kamiokande:2021jaq,Zhang:2013tua}. Enrichment with gadolinium (Gd) has significantly enhanced SK's sensitivity, potentially leading to the possibility of detection of the DSNB within the next few years~\cite{Harada:2023apz,Beacom:2003nk,Simpson:2018snj}. Furthermore, other future experiments such as Theia~\cite{Theia:2019non}, as well as those relying on coherent elastic neutrino-nucleus scattering for detection~\cite{Pattavina:2020cqc,Suliga:2021hek,Baum:2022wfc}, also exhibit promising capabilities for DSNB detection. The anticipated discovery of the DSNB also facilitates exploration of new physics beyond the SM~\cite{Moller:2018kpn, DeGouvea:2020ang,Ivanez-Ballesteros:2023lob,Bell:2022ycf,deGouvea:2022dtw,Das:2021lcr, Tabrizi:2020vmo, Das:2022xsz, Martinez-Mirave:2024hfd}.

Previously, DSNB-boosted DM was analysed by some of the present authors while focusing on scenarios where DM interacts with leptons through a heavy mediator~\cite{Das:2021lcr}. Here, we instead consider DM interactions mediated by a scalar or a vector boson of a varied mass. This not only broadens the scope of considered DM models but has crucial implications for the resulting constraints, as the energy dependence of the cross-section changes drastically for a light mediator compared to a heavy one. As we shall demonstrate, our results have significant qualitative dependence on the considered mediator mass. Further, our analysis enables the translation of the existing experimental constraints into new constraints for different DM models.

Furthermore, the interactions with SM constituents lead to in-medium attenuation of the boosted DM flux incoming towards terrestrial experiments (see e.g.~\cite{Hu:2016xas,Bringmann:2018cvk,Ema:2018bih,Dent:2019krz,Plestid:2020kdm,ArguellesDelgado:2021lek,Arguelles:2022fqq,Cappiello:2023hza} within the context of other scenarios).
We improve upon results of previous DSNB-boosted DM study~\cite{Das:2021lcr}, by some of the present authors, by including the effects of DM flux attenuation due to interactions with the electrons in the atmosphere and the Earth, using a full numerical treatment. A step in this direction was recently taken in Ref.~\cite{DeRomeri:2023ytt}, considering a simplified analytical treatment of attenuation. However, the simplifying assumptions used to obtain results from such an analytical approach are not valid when the boosted DM has kinetic energy comparable to the electron mass, particularly relevant for DM with low masses. Instead, in this work, we relax this assumption and numerically solve for the mean energy loss effects due to attenuation in the approximation of negligible DM deflection and discuss the impact on boosted DM flux.

Focusing on resulting signatures within large underground direct DM detection experiments\footnote{We note that analysis within the context of neutrino experiments is also of interest (see e.g.~\cite{Cappiello:2019qsw}).}, such as XENONnT~\cite{XENON:2023cxc}, LUX-ZEPLIN (LZ)~\cite{LZ:2022ysc} and PandaX-4T~\cite{PandaX:2022ood}, we show that attenuation plays an important role for a broad parameter space of interest in the DM mass versus cross-section plane.
We find that due to attenuation, the constraints change significantly when the full energy dependence of scattering cross-sections is taken into account. The primary impact of attenuation in the Earth's crust is the downgrading of the energy of traversing DM particles. Attenuation effects are especially pronounced when the mediator is light, resulting in a shift of the resulting constraints from terrestrial experiments towards lower DM masses.

The paper is organized as follows. In Sec.~\ref{sec:dsnb} and Sec.~\ref{sec:snbdm} we describe the DSNB flux and DM-lepton interaction models, respectively. In Sec.~\ref{sec:attenuation}, we discuss the upscattering and attenuation of DM and our numerical method to compute it. In Sec.~\ref{sec:results} we discuss our results. We conclude in Sec.~\ref{sec:conclusions}.

%%%%%%%%%%%%
%%%%%%%%%%
\section{Diffuse Supernova Neutrino Background}\label{sec:dsnb}
%%%%%%%%%%%
%%%%%%%%%%

CCSNe are among the most energetic and extreme violent events in our Universe. At any given time, the rate of CCSNe in the whole observable Universe can be as large as one per second. It is well established that in a single such explosive transient event, almost $\sim99\%$ of the binding energy of the progenitor star is carried away by $\sim 10^{58}$ emitted neutrinos.  
Accumulation of these neutrinos from historic CCSNe forms a persistent DSNB flux. The DSNB is isotropic and is composed of $\mathcal{O}(10)\,{\rm MeV}$ energy neutrinos that are as ubiquitous as neutrinos comprising the cosmic neutrino background that is a relic of the Big Bang in the early Universe, only significantly more energetic.

\begin{figure}[!t]
    \centering
    \includegraphics[width=0.6\textwidth]{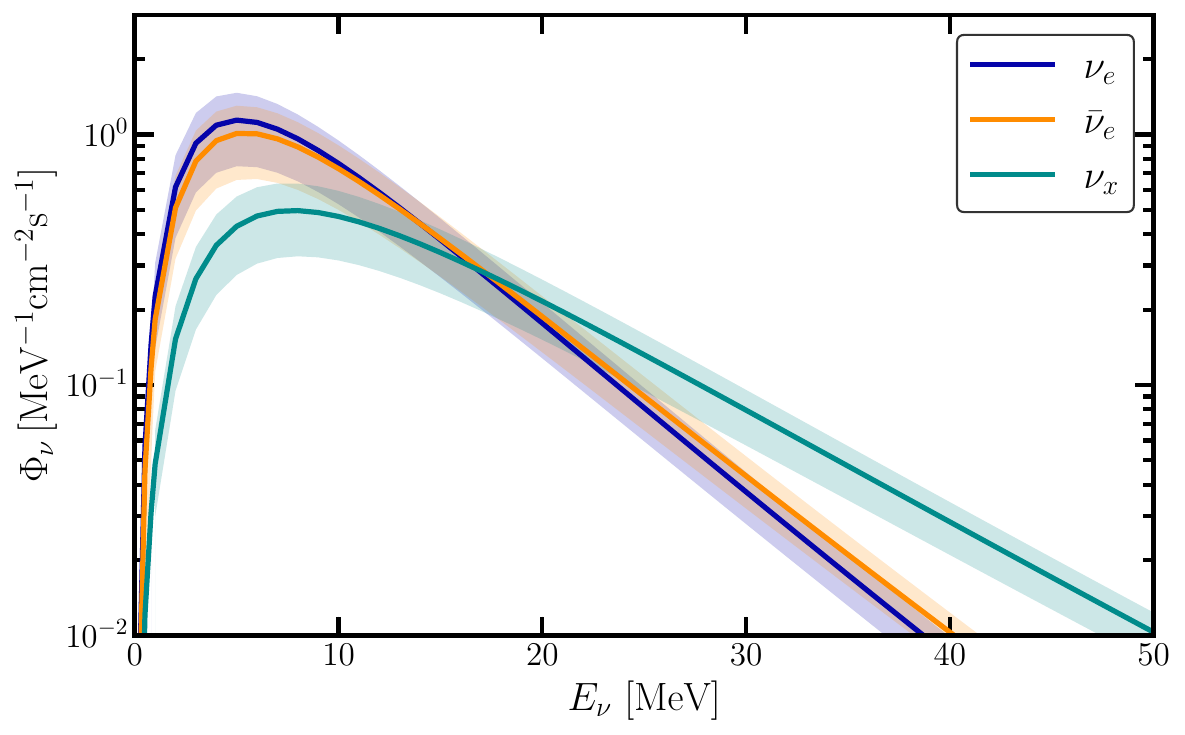}
    \caption{The diffuse supernova neutrino background (DSNB) spectra for $\nu_e$ (blue), $\Bar{\nu}_e$ (orange), and $\nu_x$ (green). The width for each flavour denotes the uncertainty from the history of the star-formation rate.}
    \label{fig:dsnb}
\end{figure}

The estimation of the DSNB spectra requires knowledge of the rate of CCSN happening in the Universe, $\Rccsn(z)$, where $z$ is the redshift of occurrence of the CCSN in the past, and the energy spectrum of the neutrinos from individual CCSN, $F_\nu(E')$. Combining these, the DSNB spectra for a given neutrino flavour can be computed as\,\cite{Horiuchi:2008jz,Beacom:2010kk}
\begin{equation}\label{eq:DSNB}
\Phi_{\nu}(E)=\int_0^{z_{\rm max}}\frac{dz}{H(z)} \Rccsn(z) F_\nu(E')|_{E'=E(1+z)}\,,
\end{equation}
where $H(z) = H_0\, \sqrt{\Omega_m (1+z)^3+\Omega_\Lambda}$ is the Hubble function with $H_0=67\,{\rm km\,s}^{-1}\,{\rm Mpc}^{-1}$, $\Omega_m=0.32$, $\Omega_\Lambda=0.68$ in the $\Lambda$CDM cosmological model\,\cite{Planck:2018vyg}. We take the maximum redshift of star formation to be $z_\mathrm{max}=6$. The neutrino emission spectrum $F(E)$ from a single CCSN can be obtained from a hydrodynamic simulation. Typically, the fluence denoting time-integrated spectra is relevant for the estimation. This is dominated by the cooling phase of the CCSN, and hence the spectrum can approximated to be a Fermi-Dirac distribution~\cite{Beacom:2010kk}
\begin{equation}\label{eq:Nuspec}
F_\nu(E)=\frac{E_\nu^{\rm tot}}{6}\,\frac{120}{7\pi^4}\,\frac{E^2}{T_\nu^4}\,\frac{1}{e^{E/T_\nu}+1}\,.
\end{equation}
Here, $E_\nu^{\rm tot}=3\times10^{53}\erg$ is the total emitted neutrino energy and we take for reference input temperatures to be $T_{\nu_e}=6.6\MeV, T_{\Bar{\nu_{e}}}=7\MeV$, and $T_{\nu_x}=10\MeV$, where $\nu_x\equiv \nu_{\mu,\tau}$\footnote{ For lower values of these temperatures, as inspired by recent studies~\cite{Newmark:2023vup}, we found that our results do not change considerably.}.
Constraints on the allowed values of the temperature have been put from null observations of the DSNB flux by SK~\cite{Super-Kamiokande:2021jaq}. Alternatively, one can also use the $\alpha-$fit parameterization of $F_\nu$, following~\cite{Keil:2002in}. We do not expect this to change the qualitative nature of our results.

The CCSN rate can be estimated from the star formation rate (SFR) $\Rsfr(z)$. To this end, we use the following fitting function from Ref.\,\cite{Kistler:2013jza},
\begin{align}
    \Rccsn(z) = \dfrac{R_{0,\mathrm{SFR}}}{143} \left[(1+z)^{-10\alpha} + \left(\dfrac{1+z}{B}\right)^{-10\beta} + \left(\dfrac{1+z}{C}\right)^{-10\gamma}\right]^{-1/10}\,.
\end{align}
Here, the parameters are as follows: $R_{0,\mathrm{SFR}}=0.0178\,\mathrm{yr^{-1}Mpc^{-3}},\, \alpha=3.4,\, \beta=-0.3,\, \gamma=-3.5$, and $B = (1+z_1)^{1-\alpha/\beta},\, C = (1+z_1)^{(\beta-\alpha)/\gamma}\times(1+z_2)^{1-\beta/\gamma}$. For a more recent measurement of the SFR, see~\cite{Ekanger:2023qzw}. We also neglect contributions from failed SNe~\cite{Moller:2018kpn}.

We consider the neutrino spectra from the SN to be processed by adiabatic Mikheyev-Smirnov-Wolfenstein flavour conversions and neglect the effects due to collective oscillations in view of the larger $\sim 40\%$ uncertainty from CCSN rate\,\cite{Horiuchi:2008jz}. We assume normal mass-ordering for neutrinos, which implies that the $\nu_e$ is mostly associated with the heaviest mass eigenstate $\nu_3$, and the $\bar{\nu}_e$ with the lightest state, $\bar{\nu}_1$. The DSNB spectra for different neutrino flavours are shown in Fig.\,\ref{fig:dsnb}, with the corresponding uncertainty from SFR. We find that the $\nu_e$ spectrum peaks at lower energy, whereas at higher energies, the dominant contribution comes from the non-electron flavour neutrinos -- $\nu_x$. Using this, we will compute the boosted DM spectra at the Earth.

%%%%%%%%%%
%%%%%%%%%%%%
\section{Dark Matter - Lepton Interactions}\label{sec:snbdm}
%%%%%%%%%%%%
%%%%%%%%%%%%
In this section, we study simple phenomenological models of DM-lepton interactions. Such interactions, responsible for boosting the DM, attenuation, and ultimately detection, can arise in a variety of models of new physics beyond SM, where the DM is coupled to the SM through some portal interactions. We consider cases where the interactions can be mediated by a vector or a scalar boson. 
%%%%%%%%%%%%%%
%%%%%%%%%%%%%%
\subsection{Vector mediated dark matter-lepton interaction}
%%%%%%%%%%%%%%
%%%%%%%%%%%%%%%
In the first example, we consider a fermionic singlet Dirac DM $\chi$ coupled to a massive vector boson $Z_{\mu}'$. The relevant interaction reads, 
\begin{equation}
\mathcal{L} \supset g_e\,\bar{e} \gamma^{\mu} e Z_{\mu}'+  g_\nu\, \bar{\nu} \gamma^\mu \nu Z_{\mu}' + g_\chi\, \bar{\chi} \gamma^\mu \chi Z_{\mu}'\,.
\end{equation}
Under the assumption $g_\nu = g_e =g$, this can be embedded in a UV-complete model. This can happen, for example, if the DM is charged under an additional $U(1)$ symmetry. We do not discuss here further details of a complete model construction. The relevant part of the  Lagrangian reads
\begin{equation}\label{eq:vectorportalmodel}
\mathcal{L} \supset \bar{\chi} \left(i \slashed{\partial} - m_\chi\right)\chi - \frac{1}{4} Z'_{\mu\nu} Z'^{\mu\nu} +\frac{1}{2}m_{Z'}^2 Z_{\mu}'Z'^{\mu}+ g \bar{L}\gamma_\mu L Z'^{\mu} + g_\chi \bar{\chi}\gamma_\mu \chi  Z'^{\mu} + \rm{h.c.} \,.
\end{equation}
Here $L$ is the SM lepton doublet.
This model permits the upscattering of $\chi$ through the following $t$-channel scattering process: $\nu(p_1)+\chi(p_2)\to \nu(p_3)+\chi(p_4)$. 
The upscattering can be captured by the following averaged matrix amplitude-squared (assuming $m_\nu=0$),
\begin{equation}\label{eq:NuChiVector}
   \vert M\vert^2 = \frac{2\,g^2  g_\chi^2}{(t-m_{Z'}^2)^2} \left[ 2 (m_\chi^2 - s)^2 + 2 st + t^2 \right]\,.
\end{equation}
The corresponding matrix-amplitude squared for direct detection with electrons is given by 
\begin{equation}\label{eq:EChiVector}
   \vert M\vert^2 = \frac{2\,g^2  g_\chi^2}{(t-m_{Z'}^2)^2} \left[2 (m_e^2 +  m_\chi^2 - s)^2 + 2 st + t^2 \right]\,.
\end{equation}
Eq.\,(\ref{eq:NuChiVector}) can be used to estimate the energy dependence of DM-neutrino differential cross-section for the vector mediator in terms of the DM kinetic energy $(T_\chi)$. In the heavy mediator limit, we find that 
\begin{equation}
\label{eq:vec1}
\frac{d\sigma_{\nu\chi}}{dT_\chi} \propto  \left\{\begin{array}{ll}
 \frac{\mchi}{m_{Z'}^4}\ , &\quad \text{for $E_\nu \gg m_\chi$}\\%(T_\chi)
\\
\frac{T_\chi}{ m_{Z'}^4} \ , &\quad \text{for $m_\chi \gg E_\nu $\,.}
\end{array}\right.
\end{equation}
On the other hand, if the mediator is light, the differential cross-section follows,
\begin{equation}
\frac{d\sigma_{\nu\chi}}{dT_\chi} \propto  \left\{\begin{array}{ll}
 \frac{1}{T_\chi^2 m_{\chi}} \ , &\quad \text{for $E_\nu \gg m_\chi$}\\
\\
\frac{1}{T_\chi m_{\chi}^2} \ , &\quad \text{for $m_\chi \gg E_\nu $\,.}
\end{array}\right.
\end{equation}
A similar dependence can be worked out for the DM-electron differential cross-section. We find, in the heavy mediator limit, 
\begin{equation}
\frac{d\sigma_{e\chi }}{dT_e} \propto  \left\{\begin{array}{ll}
 \frac{m_e}{m_{Z'}^4} \ , &\quad \text{for $T_e \gg m_\chi$}\\
\\
\frac{m_\chi m_e }{T_\chi m_{Z'}^4} \ , &\quad \text{for $m_\chi \gg T_e $\,.}
\end{array}\right.
\end{equation}
%%%
On the other hand, if the mediator is light, we find that
\begin{equation}
\label{eq:vec2}
\frac{d\sigma_{e\chi}}{dT_e} \propto  \left\{\begin{array}{ll}
 \frac{1}{ m_e T_\chi^2} \ , &\quad \text{for $T_e  \gg m_\chi$}\\
\\
\frac{(m_\chi + m_e)^4}{T_\chi^3 m_e^3 m_\chi} \ , &\quad \text{for $m_\chi \gg T_e $,}
\end{array}\right.
\end{equation}
where $T_e$ is the electron recoil kinetic energy.

%%%%%%%%%%%%%%
%%%%%%%%%%%%%%
\subsection{Scalar mediated dark matter-lepton interaction}
%%%%%%%%%%%%%%
%%%%%%%%%%%%%%%
The second example considers the DM to be a SM singlet Dirac fermion, $\chi$, which can interact with neutrinos and electrons through two scalar mediators, $\Phi^0$ and $\Phi^-$. This proceeds through the interaction,
\begin{equation}
\mathcal{L} \supset g_\nu \bar{\nu} \chi \Phi^0 - g_e \bar{e} \chi \Phi^- \,.
\end{equation}
Just like the vector model, following $g_\nu = g_e=g$, this can be embedded in a UV-complete model.
The SM can be augmented with a new scalar doublet, $\Phi \in \left[\Phi^-, \Phi^0\right]$. The corresponding Lagrangian can be written as
\begin{equation}\label{eq:scalarportalmodel}
\mathcal{L} \supset   \bar{\chi} \left(i \slashed{\partial} - m_\chi\right)\chi + \partial_\mu \Phi^\dagger\partial^\mu \Phi + g \bar{L}\Phi\chi - \mu_\phi |\Phi|^2 - \lambda_\Phi |\Phi|^4 - \lambda_{\Phi H} |H|^2|\Phi|^2 + \rm{h.c.} \,,
\end{equation}
where $H$ is the SM Higgs field and $L$ is the SM lepton doublet. The assumption of a dark sector symmetry for $\chi$ and $\Phi$ can allow one to set the interaction term $\bar{L}H\chi$ to zero. For our analysis, we consider scenarios, $m_\chi < m_\Phi$, where $m_\phi^2 = \mu_\Phi^2 + \lambda_{\Phi H} v_{\rm EW}^2$.

The matrix amplitude squared for the upscattering of $\chi$ can be computed as
\begin{equation}\label{eq:NuChiScalar}
    \vert M\vert^2 = \frac{g^4}{2} \frac{(t - m_\chi^2)^2}{(t-m_\Phi^2)^2}\,.
\end{equation}
The same interaction also allows the direct detection of boosted DM through the electron scattering channel. The matrix amplitude-squared for the process is given by
\begin{equation}\label{eq:eChiScalar}
    \vert M\vert^2 = g^4 \frac{\left[(m_\chi+m_e)^2-t\right]^2}{(t-m_\Phi^2)^2}\,.
\end{equation}
Note that in the case of $\chi-e$ scattering, a $t$-channel resonance can take place if $t=m_\phi^2$ is satisfied. This $t$-channel singularity is a generic feature of scattering processes where particles are exchanged at the vertices. This can be interpreted in terms of the stability of one of the particles in the initial and final states~\cite{Grzadkowski:2021kgi}. In our study, we will not worry about such singularities as it happens for that region of the parameter space which allows for the electron to decay and hence is ruled out from observations. No such singularities exist for the $\chi-\nu$ channel due to the requirement of $m_\phi > m_\chi$.

Eq.\,(\ref{eq:NuChiScalar}) can be used to estimate the energy dependence of DM-neutrino differential cross-section in terms of the DM kinetic energy $(T_\chi)$, relevant for our analysis. In the heavy mediator limit, we find that 
\begin{equation}
\label{eq:sca1}
\frac{d\sigma_{\nu\chi}}{dT_\chi} \propto  \left\{\begin{array}{ll}
 \frac{m_\chi}{m_{\Phi}^4} \ , &\quad \text{for $E_\nu \gg m_\chi$}\\
\\
\frac{\mchi}{ m_{\Phi}^4}\left(1-\frac{4\mchi T_\chi}{m_\Phi^2}\right) \ , &\quad \text{for $m_\chi \gg E_\nu $\,.}
\end{array}\right.
\end{equation}
On the other hand, if the mediator is light, the differential cross-section follows,
\begin{equation}
\frac{d\sigma_{\nu\chi}}{dT_\chi} \propto  \left\{\begin{array}{ll}
 \frac{1}{T_\chi^2 m_{\chi}} \ , &\quad \text{for $E_\nu \gg m_\chi$}\\
\\
\frac{1}{m_\chi^3} \ , &\quad \text{for $m_\chi \gg E_\nu $\,.}
\end{array}\right.
\end{equation}
A similar dependence can be worked out for the DM-electron differential cross-section. We find, in the heavy mediator limit, 
\begin{equation}
\frac{d\sigma_{e\chi }}{dT_e} \propto  \left\{\begin{array}{ll}
 \frac{m_e}{m_\Phi^4}\ , &\quad \text{for $T_e  \gg m_\chi$}\\
\\
\frac{\mchi m_e}{T_\chi m_\Phi^4}\ , &\quad \text{for $m_\chi \gg T_e $\,.}
\end{array}\right.
\label{eq:approx_echi_scalar}
\end{equation}
%%%
On the other hand, if the mediator is light, we find that
\begin{equation}
\label{eq:sca2}
\frac{d\sigma_{e\chi}}{dT_e} \propto  \left\{\begin{array}{ll}
 \frac{1}{ m_e (2T_\chi-m_e)^2} \ , &\quad \text{for $T_e  \gg m_\chi$}\\
\\
\frac{\mchi m_e}{T_\chi(\mchi-m_e)^4}\ , &\quad \text{for $m_\chi \gg T_e $,}
\end{array}\right.
\end{equation}
where $T_e$ is the electron kinetic energy.
\begin{figure}[!t]
    \centering
    \includegraphics[width=0.45\textwidth]{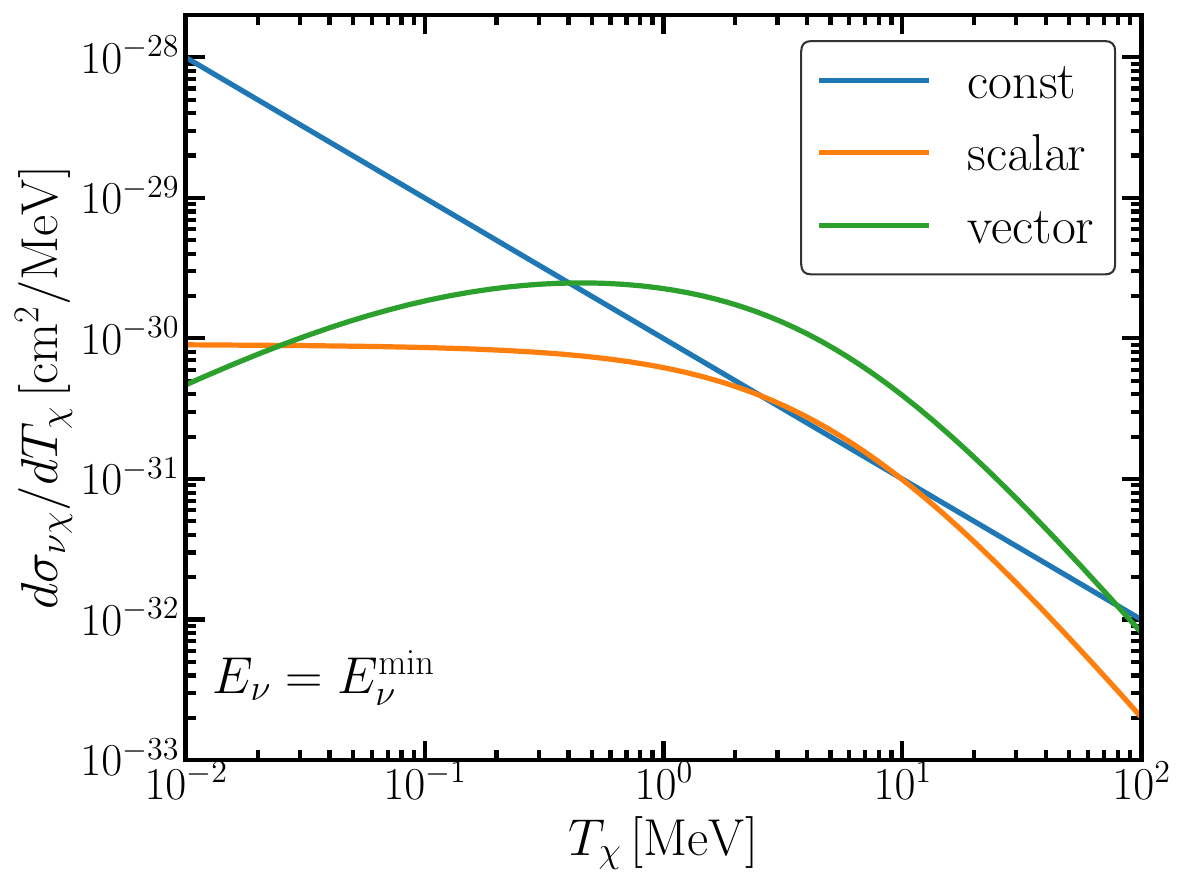} 
    \includegraphics[width=0.45\textwidth]{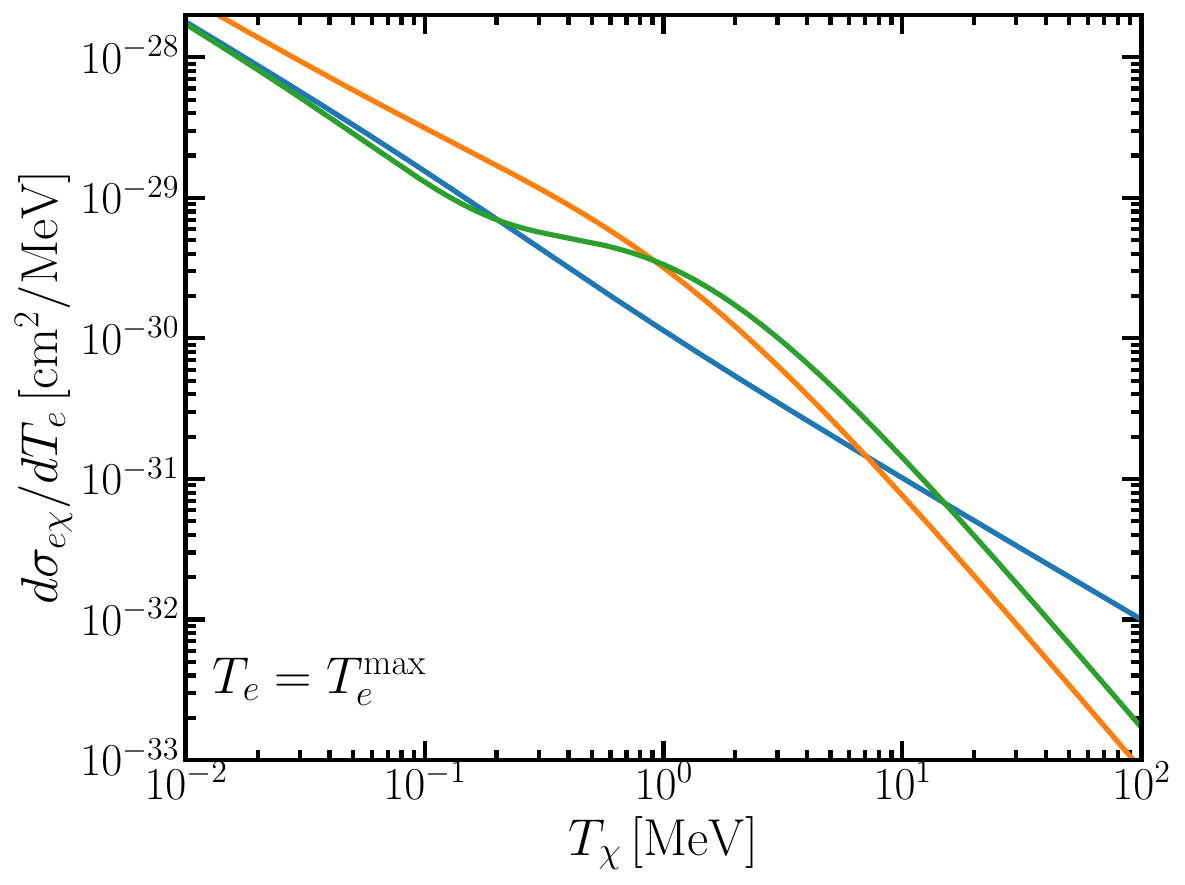}
    \includegraphics[width=0.45\textwidth]{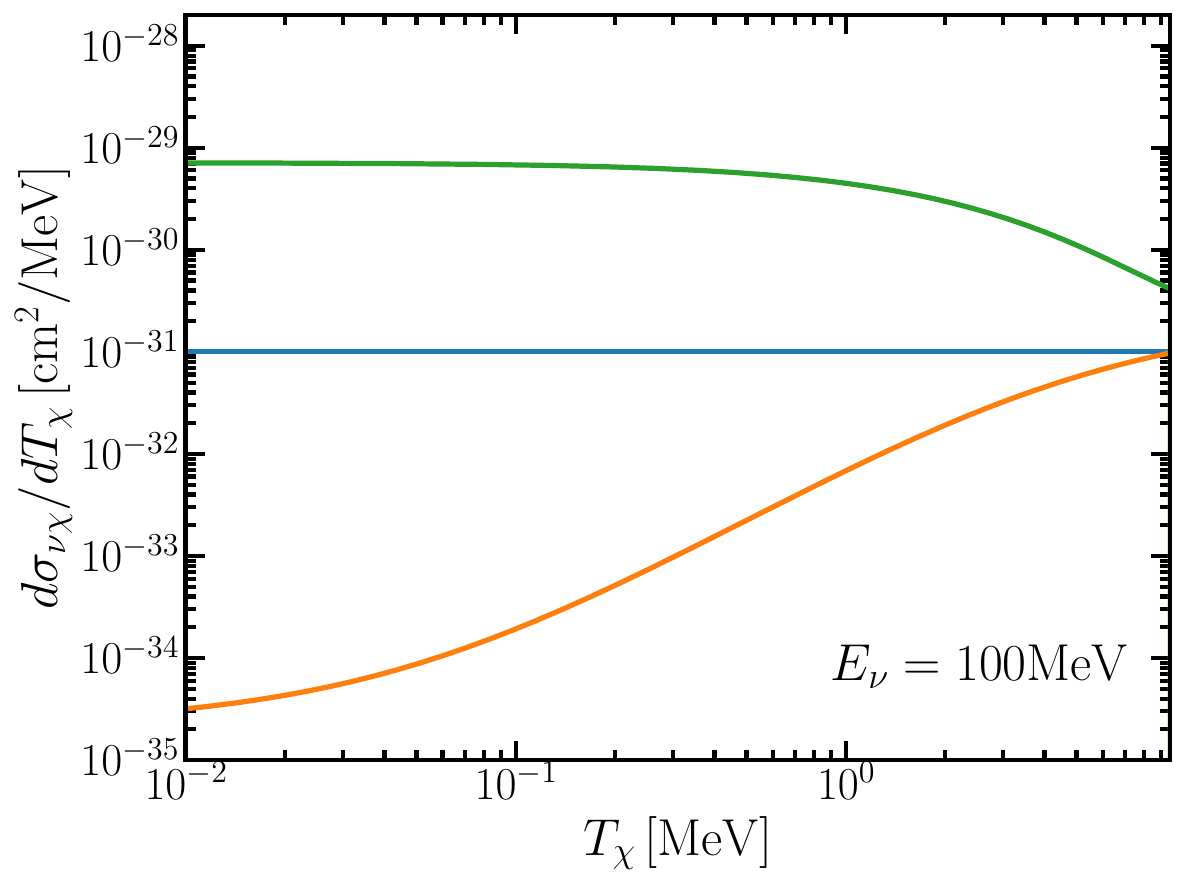} 
    \includegraphics[width=0.45\textwidth]{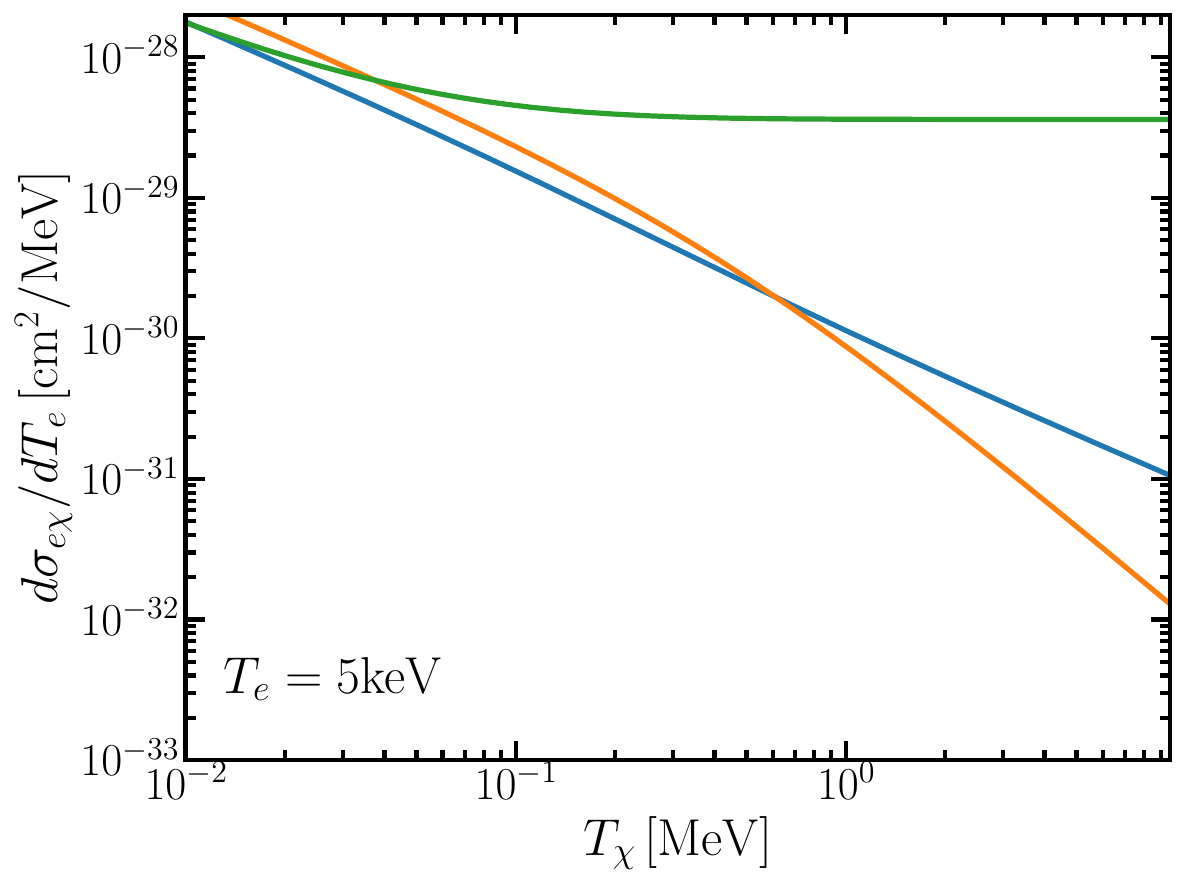}
    \caption{Comparison of the differential cross-section of neutrinos and electrons with DM. We present benchmark point with a  mediator mass of $1\MeV$, $m_\chi = 0.1\text{MeV}$ and fix the scalar and gauge couplings such that $\bar{\sigma}_{e\chi} = 10^{-30}\,\text{cm}^2$.
    Top Left: Differential cross-section for the interaction of neutrinos with DM at rest as a function of kinetic energy gained by DM. The neutrino energy is fixed to the minimum energy required to boost DM to kinetic energies $T_\chi$.
    Top Right: Differential cross-section for scattering of BDM with electrons at rest as a function of BDM kinetic energy. We fix the recoil electron energy to the maximum value that can be achieved upon scattering with BDM of energy $T_\chi$.
    Bottom row: Same as above but with the incident neutrino energy and electron recoil energy fixed to $E_\nu = 10\, \mathrm{MeV}$ and $T_e = 5\,\mathrm{keV}$ respectively.
    }
    \label{fig:attenuation_example_2}
\end{figure}

These dependencies discussed above can be confirmed from Fig.\,\ref{fig:attenuation_example_2}, which compares the differential cross-section as a function of $T_\chi$ between the constant cross-section case and the case with scalar/vector mediators. The left panels focus on DM-neutrino cross-sections, whereas the right panels show DM-electron cross-sections. Previous studies with constant DM cross-sections usually assumed these two cross-sections to be identical. However, it is important to stress that these two cross-sections are different in general, and there is no reason for them to be considered equal unless the underlying model demands it to be. The top and bottom panels are for two different representative values of $E_\nu$ and $T_e$ respectively. For the top left panel, the neutrino energy is fixed to the minimum energy required to boost DM to kinetic energies $T_\chi$, whereas the bottom left panel shows the plots for a representative value $E_\nu =10\,{\rm MeV}$. Similarly, for the top right panel, we fix the recoil electron energy $(T_e)$ to the maximum value that can be achieved upon scattering with BDM of energy $T_\chi$, whereas for the bottom right panel, we choose $T_e=5\,{\rm keV}$. The limiting behaviour of the differential cross-sections can be qualitatively understood from Eqs.\,(\ref{eq:vec1})-(\ref{eq:vec2}), and Eqs.\,(\ref{eq:sca1})-(\ref{eq:sca2}) respectively.

%%%%%%%%%%%%%
%%%%%%%%%%%%%
\section{Boosted Dark Matter}\label{sec:attenuation}
%%%%%%%%%%%%
%%%%%%%%%%%%
The neutrinos from the DSNB can scatter off ambient DM particles and transfer a part of their energy, resulting in a fraction of the DM being boosted to high energies. As a result, the DM particles have kinetic energy that significantly exceeds the energy from their virial motion in galactic structures. From kinematic considerations, it follows that for a neutrino hitting a DM particle at rest, the transferred kinetic energy is given by
\begin{equation}
\begin{aligned}
    T_\chi &= T_\chi^{\text{max}} \left(\frac{1-\cos\theta}{2}\right)\,,\qquad
    T_\chi^{\text{max}} &= \frac{E_\nu^2}{E_\nu + m_\chi/2}\,.
\end{aligned}
\end{equation}
Here $E_\nu$ is the neutrino energy and, $\theta$ is the scattering angle in the centre of momentum frame.
The resulting BDM flux in the Milky Way halo is obtained by performing a line-of-sight integral over all possible directions of incoming DM that was upscattered by the DSNB. For a given DSNB flux spectrum $d\Phi_\nu/dE_\nu$, the flux of boosted DM reads \cite{Das:2021lcr,Bardhan:2022bdg,Bringmann:2018cvk,DeRomeri:2023ytt}
\begin{equation}
    \frac{d\Phi_\chi}{dT_\chi} = \int \frac{d\Omega}{4\pi}\int_{\text{l.o.s.}} dl \int_{E_\nu^{\text{min}}}^{E_\nu^{\text{max}}} dE_\nu \frac{\rho_\chi(l)}{m_\chi} \frac{d\Phi_\nu}{dE_\nu} \frac{d\sigma_{\nu\chi}}{dT_\chi} \equiv D_{\text{halo}} \int_{E_\nu^{\text{min}}}^{E_\nu^{\text{max}}} dE_\nu \frac{1}{m_\chi} \frac{d\Phi_\nu}{dE_\nu} \frac{d\sigma_{\nu\chi}}{dT_\chi} \,,
    \label{eq:upscatterd_flux}
\end{equation}
In the above equation, we have exploited the factorization of halo dependence and the underlying particle physics. We assume that the DM density in the MW halo $\rho_\chi (r) $ follows a Navarro-Frenk-White (NFW) profile\,\cite{Navarro_1996}
\begin{equation}
    \rho_\chi (r) = \rho_{s}\frac{ \left(\frac{r}{r_s}\right)^{-\gamma}} {1+\left(\frac{r}{r_s}\right)^{\beta -\gamma} }\,,
\end{equation}
where $r$ is the radial distance from the Galactic centre (GC),
and $\beta,\gamma,r_s$ are fitting parameters. Following the NFW profile $\left(\beta,\gamma,r_s\right)= \left(3,1,20\,{\rm kpc} \right)$, 
we get $D_\text{halo} = 2.04\times 10^{25}\,\text{MeV cm}^{-2}$. The precise value of $D_\text{halo}$ depends on assumptions made on the nature of the DM halo profile. For various choices of the halo following~\cite{Ng:2013xha}, we find that the difference is not more than $\mathcal{O}(1\%)$, which agrees with what was found in~\cite{DeRomeri:2023ytt}. 

%%%%%%%%%%
\subsection{Attenuation}
%%%%%%%%%
On their way to the detector, boosted DM particles can interact with electrons in the atmosphere and the earth and lose energy before reaching the detector site. For sufficiently strong interactions, this can lead to distortion and ultimately attenuation of the differential BDM flux. We model the mean energy loss of a single DM particle travelling through a medium by the energy loss equation (e.g.~\cite{Bringmann:2018cvk,Ema:2018bih,Dent:2019krz})
\begin{equation}
    \frac{dT_\chi}{dx}(x) = - \sum_{i} n_i(x) \int_{0}^{T_i^{\text{max}}} dT_i \, T_i \frac{d\sigma_{i\chi}}{dT_i}\,.
    \label{eq:energyloss}
\end{equation}
Here we sum over all medium constituents that may participate in the scattering, their respective number densities $(n_i)$ as a function of distance $x$, and integrate over the energy lost in a single interaction. Note that only in the case of elastic scattering, $T_i$ correspond to the recoil energy. In this case, the maximum energy that can be transferred is \cite{Bringmann:2018cvk}
\begin{equation}
    T_i^{\text{max}} = \frac{T_\chi^2 + 2m_\chi T_\chi}{T_\chi + (m_\chi+m_i)^2/(2m_\chi)}\,,
\end{equation}
where $m_i$ is the mass of the target particle, assumed to be at rest. 

Under certain assumptions, Eq.(\ref{eq:energyloss}) admits an analytic solution. For $d\sigma_{i\chi}/{dT_i} = \sigma_i/T_i^{\text{max}}$ - the constant cross-section assumption, the integral can be performed analytically. If we also assume $T_\chi \ll m_i$, we find
\begin{equation}
    \frac{dT_\chi}{dx} = -\frac{1}{2} \sum_i n_i \sigma_i T_i^{\text{max}} \approx - \frac{T_\chi^2+2m_\chi T_\chi}{2m_\chi l}\,, \quad \text{where} \quad  l^{-1} = \sum_i n_i \sigma_i \frac{2m_i m_\chi}{(m_i+m_\chi)^2}\,.
\end{equation}
Solving the resulting ODE yields
\begin{equation}
    T_\chi(x) \equiv T_\chi^x = T_\chi^0 \frac{e^{-x/l}}{1+\frac{T_\chi^0}{2m_\chi}\left(1-e^{-x/l}\right)}\,,
\end{equation}
which can also be inverted to find $T_\chi^0(T_\chi^x, x)$. This allows us to express the attenuated flux at the detector as
\begin{equation}
    \frac{d\Phi_\chi}{dT_\chi^x} = \int \frac{d\Omega}{4\pi}\,\frac{d\Phi_\chi^0}{dT_\chi^0}\, \frac{dT_\chi^0}{dT_\chi^x}\,,
    \label{eq:detector_flux}
\end{equation}
where we take into account that BDM flux from different directions travelled a different distance $x$ through the earth, depending on the angle $\theta$ of arrival to the detector.
\cite{Bringmann:2018cvk, DeRomeri:2023ytt}

The analytic solution is however not suitable for our case for multiple reasons. In the following, we only consider scattering on electrons and assume a constant electron density $n_e = 8\times 10^{23}\,\text{cm}^{-3}$ \cite{DeRomeri:2023ytt}.  For typical BDM energies, we do not expect $T_\chi \ll  m_e$ to be valid all the time. Moreover, the assumption of a constant cross-section $\left(d\sigma_{i\chi}/{dT_i} = \sigma_i/T_i^{\text{max}}\right)$ is an oversimplification. Depending on the details of the interaction, the probability, and hence the efficiency of mean energy loss, depends on the current mean energy $T_\chi^x$. Thus, different parts of the BDM energy spectrum experience a different energy loss rate, which can ultimately lead to an over or underestimation of detector events compared to the constant cross-section case. This necessitates a numerical solution of  Eq.\,(\ref{eq:energyloss}). We describe the procedure below.

The full numerical implementation of the upscattering, boosting, and arrival at the detector follows an instructive sequence of steps. After selecting a specific model, we specify all relevant parameters, e.g., DM mass, mediator mass, couplings, etc. 
We use  Eq.\,(\ref{eq:upscatterd_flux}) for a given model to numerically find the unattenuated flux of BDM at earth. To find the attenuated BDM at some detector depth $h_d$, we repeatedly solve Eq.\,(\ref{eq:energyloss}) for many different initial energies $T_0^\chi$. We find a grid of $T_\chi^x(x, T_\chi^0)$ for every point in the parameter space on which we can smoothly interpolate. For any fixed value of $x$, we can determine $T_0^\chi(T_\chi^x)$ and the Jacobian $\frac{dT_\chi^x}{dT_\chi^0}(T_\chi^0(T_\chi^x, x), x)$ and exploit the relation $\frac{dT_\chi^x}{dT_\chi^0} =  \left(\frac{dT_\chi^0}{dT_\chi^x}\right)^{-1}$. The differential flux at the detector site is now given by Eq.\,(\ref{eq:detector_flux}), which can also be turned into an integration over $x$. For the relation of the angle $\theta$ to the overburden $x$, see \cite{DeRomeri:2023ytt}. 

The numerical scheme as described above comes with several technical difficulties. To connect the entire BDM spectrum at the surface to that at the detector site, Eq.\,(\ref{eq:energyloss}) needs to be solved repeatedly even for a single point in parameter space to obtain a reliable grid for interpolation. This requires a stable ODE solution over orders of magnitude in depth and energy for many initial energies which also vary over the dominant range of energies for the BDM flux at the surface. Depending on model parameters, the rate of energy loss also varies by orders of magnitude in a parameter scan for which we have to guarantee the stability of the solver. Moreover, the solution exhibits singular behaviour, i.e., full loss of energy within finite depth. The final result, by virtue of the angular integration in Eq.\,(\ref{eq:detector_flux}), depends on very different energy loss regimes and combines singular and non-singular partial solutions.

Although such a solution is expected to be a significant improvement on previous treatments, the resulting BDM fluxes and derived constraints provide only a conservative estimate. Here we work with the assumption of negligible deflection of BDM particles. Our treatment also neglects multiple scatterings which become important as $\lambda_\text{scat} = (n\sigma)^{-1}$ no longer exceeds the depth of the detector location $h_d$ by a significant margin.

%%%%%%%%%%
%%%%%%%%%%%%
\section{Results}\label{sec:results}
%%%%%%%%%%
%%%%%%%%%%
For all the following analyses, we set the relevant couplings to be the same, i.e.,  $g_\nu= g_e = g_\chi=g$. We will comment later on the possibility of relaxing this assumption. 
To compare with results from the direct detection experiments, we use
the effective cross-section
\begin{equation}
    \bar{\sigma}_{e\chi} = \frac{g^4}{\pi} \frac{\mu_{e\chi}^2}{(q_\text{ref}^2+m_\text{med}^2)^2}\,,
    \label{eq:sigmadef}
\end{equation}
which can be used to replace couplings in the differential cross-section with the effective cross-section. We follow the conventional definition of $q_\text{ref} = \alpha m_e$, involving the fine structure constant, to compare with other works. Note that for a variety of models, this gives the correct scale of momentum transfer, for e.g., if we consider an energy-independent cross-section or a vector mediator. However, the scalar model we consider is more subtle. Due to the conversion of particle species by a t-channel exchange, there is a mass splitting of $\Delta m$ between the incoming and outgoing particles on the same fermion line. Thus, a second mass scale is present in the process. The momentum transfer to overcome this gap may be much larger than the typical transfer naively expected from $q_\text{ref} = \alpha m_e$. In those circumstances, $q_\text{ref} = \Delta m$ is a more natural definition. However to maintain comparability with previous results we adopt the original definition of $q_\text{ref}$ exclusively. In some cases, this leads to an apparent offset of the scalar cross-section when compared to the constant one. This is a matter of definition and not connected to any physical effects.

%%%%%%%%%
%%%%%%%%%%
\subsection{Impact of Lorentz structure on upscattering and attenuation}
%%%%%%%%%
%%%%%%%%%
\begin{figure}[!t]
    \centering
    \includegraphics[width=0.49\textwidth]{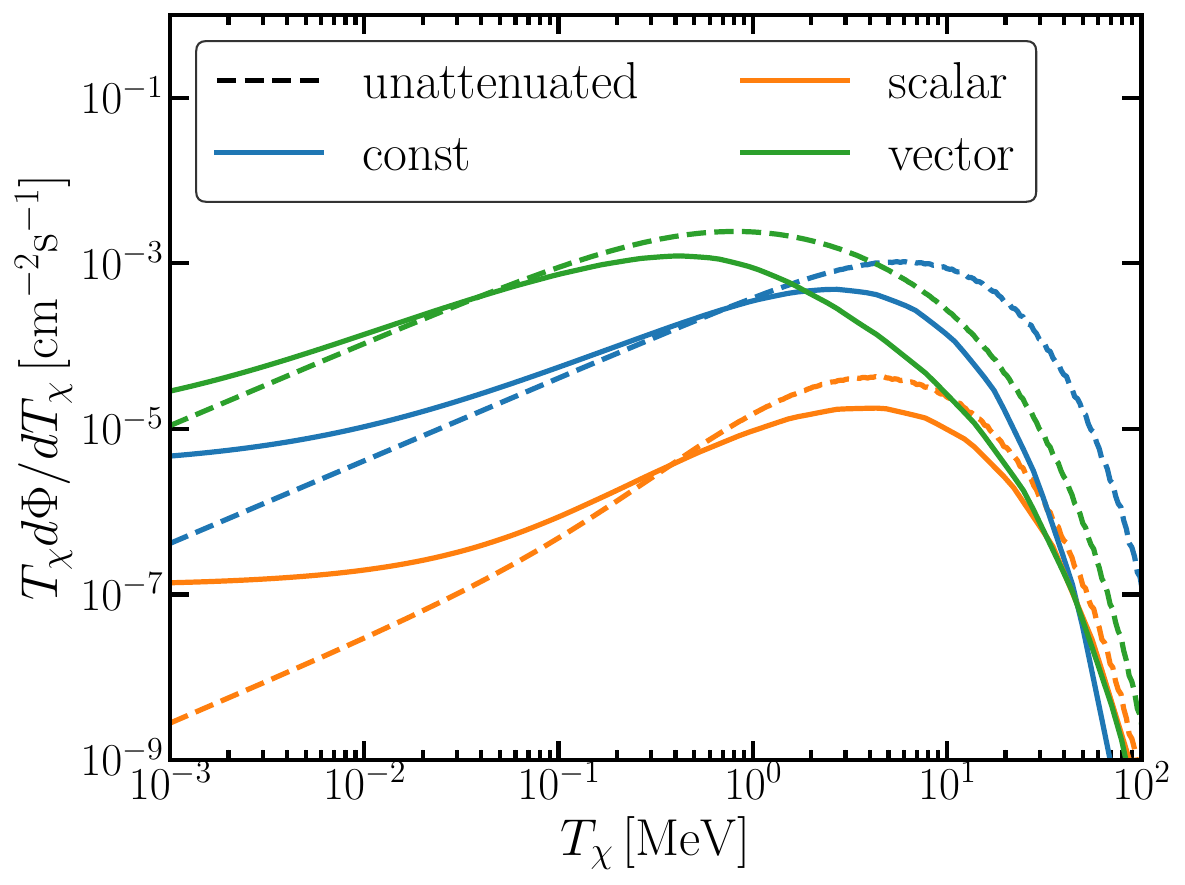}    
    \includegraphics[width=0.49\textwidth]{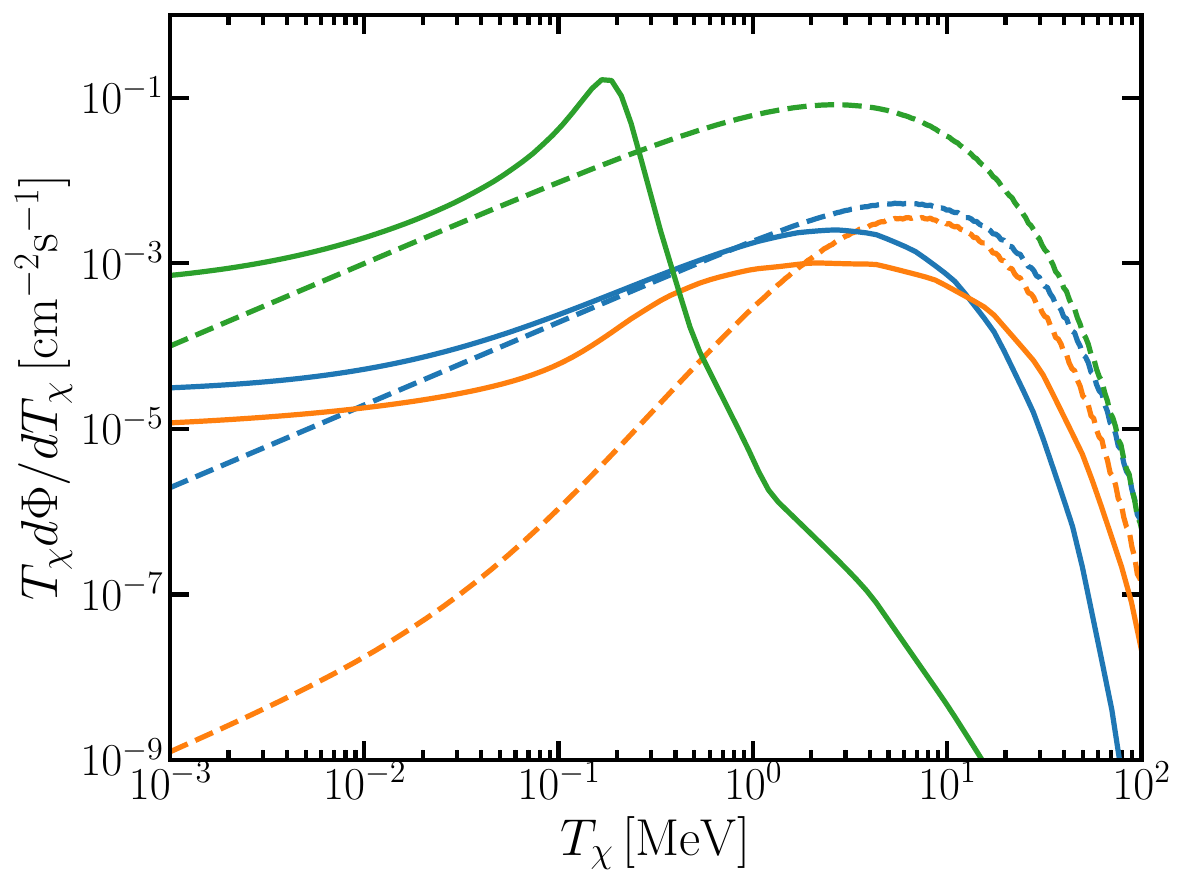}
    \caption{Attenuated and unattenuated BDM spectra for two benchmark points for the models we consider. For the left plot we assume $\bar{\sigma}_{e\chi} = 10^{-30}\,\text{cm}^2$, mediator $1\MeV$, and $m_\chi = 0.5\MeV$. The right plot shows the spectra for $\bar{\sigma}_{e\chi} = 10^{-30}\,\text{cm}^2$, mediator $1\MeV$, and $m_\chi = 0.1\MeV$. To calculate the attenuated spectrum we employ a detector location at depth $h_d = 1.4\,\mathrm{km}$. The definition of $\bar{\sigma}_{e\chi}$ is given in Eq.\,(\ref{eq:sigmadef}).
    }
    \label{fig:attenuation}
\end{figure}

We illustrate the effect of Lorentz structure on DM upscattering and attenuation of the DM flux in Fig.\,\ref{fig:attenuation} for two different benchmark points.
The resulting BDM spectra are best understood in light of the previous discussion on the energy dependence of the cross-sections. Compared to the constant cross-section case (dashed blue), the inclusion of the underlying Lorentz structure can change the BDM spectra considerably. In particular, upscattering by neutrinos is suppressed for the scalar model (dashed green) and slightly enhanced for the vector model (dashed orange) for the same effective cross-section, as compared to the constant case. This can be seen in the unattenuated BDM spectra for both benchmark points. 

Similarly, the energy dependence of DM electron scattering has a strong impact on the attenuation of the BDM flux arriving at the detector, as shown by the solid coloured lines in Fig.\,\ref{fig:attenuation}. This is due to the behaviour of the energy-dependent cross-sections, as shown in Fig.\,\ref{fig:attenuation_example_2}  and discussed in section \ref{sec:snbdm}. For example, we find that for the benchmark point, $\bar{\sigma}_{e\chi} = 10^{-30}\,\text{cm}^2$, mediator $1\,\MeV$, and $m_\chi = 0.1\,\MeV$, we find a peak followed by a strong falloff in the BDM spectra for the vector model. This is because, for these values of the parameters, the differential $e-\chi$ spectra remain almost constant with $T_\chi$ as shown in the bottom right panel of Fig.\,\ref{fig:attenuation_example_2}. As a result, depending on the energy of the BDM, attenuation affects the spectra in a qualitatively different manner, thereby leading to model-dependent BDM fluxes at the detector sites, which cannot be accurately approximated by a constant cross-section. The variation in the final signal event rate in the detector for different models can also be described using similar arguments.

%%%%%%%%%%%%
\subsection{Expected electron recoil event rates}
\begin{figure}[!t]
    \centering   
    \includegraphics[width=0.32\textwidth]{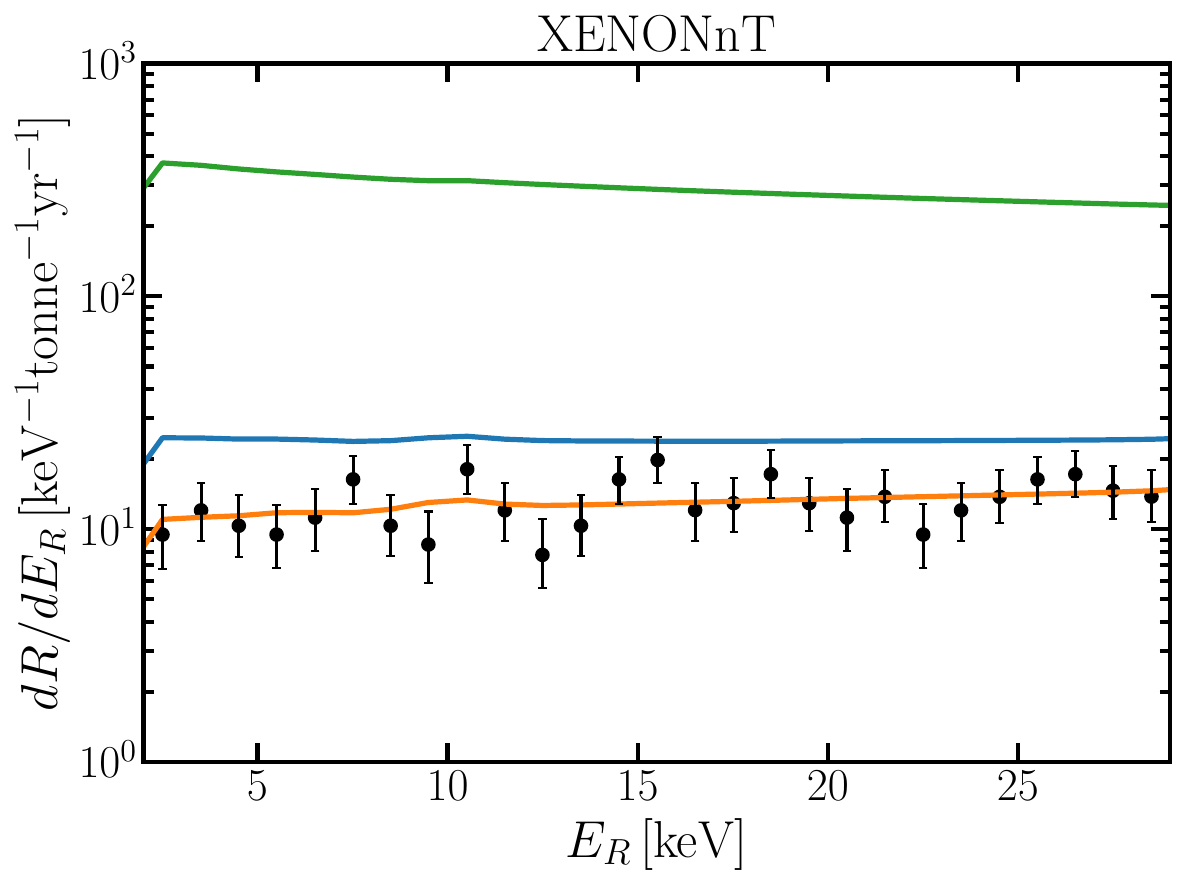}
    \includegraphics[width=0.32\textwidth]{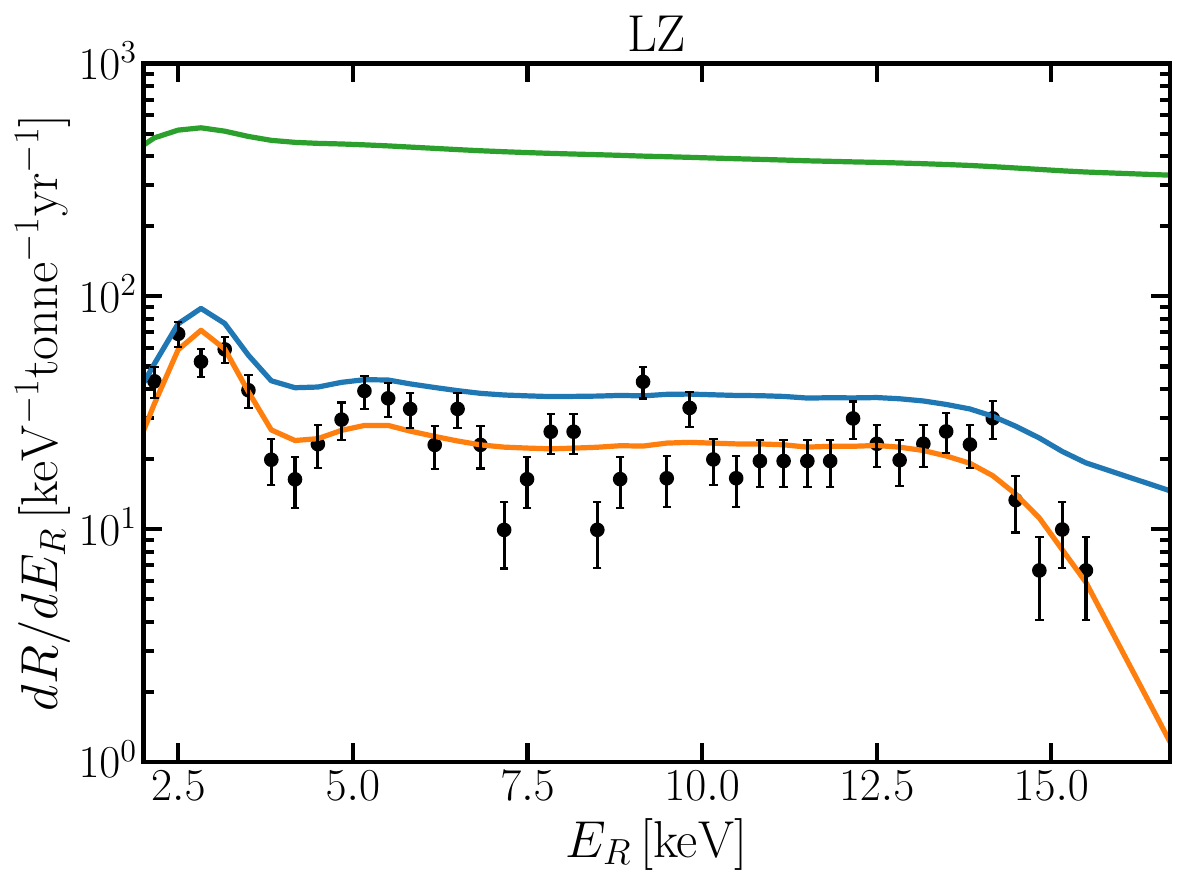}
    \includegraphics[width=0.32\textwidth]{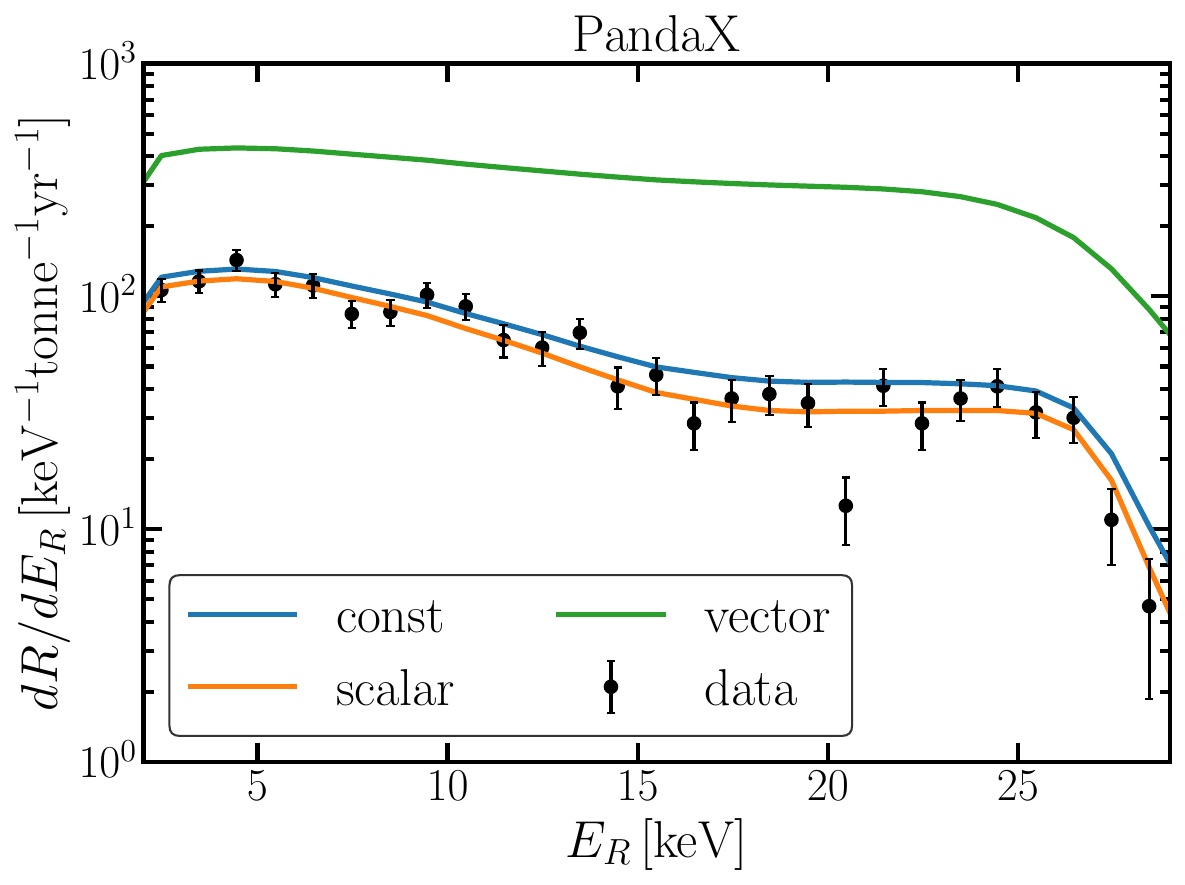}
    \caption{
    Expected electron recoil event rates in XENONnT (\emph{left}), LZ (\emph{middle}), and PandaX (\emph{right}) with respective experimental background for benchmark point $\bar{\sigma}_{e\chi} = 10^{-30}\,\text{cm}^2$, mediator mass $1\MeV$ and $m_\chi = 0.5\MeV$. We show the rates for constant cross-section (blue), vector (green), and scalar (orange) mediator models.
    }
    \label{fig:rates}
\end{figure}

%%%%%%%%%%%%
The differential electron recoil event rate is given by
\begin{equation}
    \frac{dR}{dT_e} = N_e \int dT_\chi \frac{d\Phi_\chi}{dT_\chi^z} \frac{d\sigma_{e\chi}}{dT_e}\,.
    \label{eq:event_rate}
\end{equation}
For the number of electron targets in the detector we find $N_e = M_\text{det}/m_\text{Xe} Z_\text{eff}(T_e)$, involving the total mass of detector material, the mass of a xenon atom and $Z_\text{eff}(T_e)$ is the general energy-dependent effective charge number seen by a recoiling DM particle. Throughout the analysis, we assume constant $Z_\text{eff}(T_e) \approx 40$. We take the convolution of the resulting recoil spectrum with a detector-specific resolution function. For all experiments considered this resolution function is Gaussian in shape
\begin{equation}
    I(E_R, T_e) = \frac{1}{\sqrt{2\pi \sigma(T_e)^2}} \exp{ \left( -\frac{\left(T_e-E_R\right)^2}{2\sigma(T_e)^2}\right) }\,,
\end{equation}
where $T_e$ is the ``true" deposited energy and $E_R$ the observed energy in the detector. We use the detector resolutions (in $\text{keV}$) provided by the respective collaboration, i.e. $\sigma_\text{XE} = 0.31 \sqrt{E/\text{keV}} + 0.0037 E/\text{keV}$\,\cite{Aprile_2020}, $\sigma_\text{LZ} = 0.323 \times 10^{-1.5} \sqrt{E/\text{keV}}$\,\cite{LZ:2022lsv}, and $\sigma_\text{PA} = 0.073 + 0.173 E/\text{keV} - 6.5\times 10^{-3} (E/\text{keV})^2 + 1.1\times 10^{-4} (E/\text{keV})^3$\,\cite{PandaX:2022ood} for XENONnT, LZ and PandaX, respectively. The convoluted signal is then multiplied by the respective efficiency function also provided by the collaboration. 
We show example electron recoil spectra from BDM scattering in three experiments for the constant cross-section case, and a benchmark point for the scalar and the vector model in Fig.\,\ref{fig:rates}, along with the data. In the next subsection, we proceed to perform a statistical analysis of the hypothesis with the experimental data.

%%%%%%%%%%%
\subsection{Statistical analysis}
%%%%%%%%%%%
To calculate the constraints on the parameter space $(\bar{\sigma}_{e\chi}, m_\chi)$, we use the following $\chi^2$ statistic, 
\begin{equation}
    \chi^2 = \sum_{E_i} \frac{\left(R_i^\text{pred}(E_i) - R_i^\text{exp}(E_i)\right)^2}{\sigma_i^2(E_i)}\,,
\end{equation}
for each of the experiments considered. 
Here $R_i^\text{pred}$ denotes the predicted event rates, which include the DSNB boosted DM contribution as well as the experimental background. For the fixed background, we use what is provided as the best-fit background model for each experiment. Thus, constraints derived from our approach are conservative, since we do not attempt a joint fit of the BDM and the background. Reported event rates from the experimental collaborations are denoted by $R_i^\text{exp}$.
The net uncertainty in the model and data is given by 
\begin{equation}
  \sigma_i^2(E_i)=  R_i^\text{pred}(E_i) + \sigma_{\text{Di}}^2(E_i)\,,
\end{equation}
where we estimate the event rate uncertainties $(\sigma_{\text{Di}})$ by combining a Poissonian counting error on the total predicted event rate with the experimental uncertainty from the experiment. For the exclusion contours, we use a $\chi^2$ difference to the best-fit background model, i.e. $\Delta \chi^2 = \chi^2 - \chi^2_\text{bkg}$ and exclude regions with $\Delta \chi^2 > 4.61$ at $90\%$ confidence level (CL).

\begin{figure}[!t]
    \centering
    \includegraphics[width=0.49\linewidth]{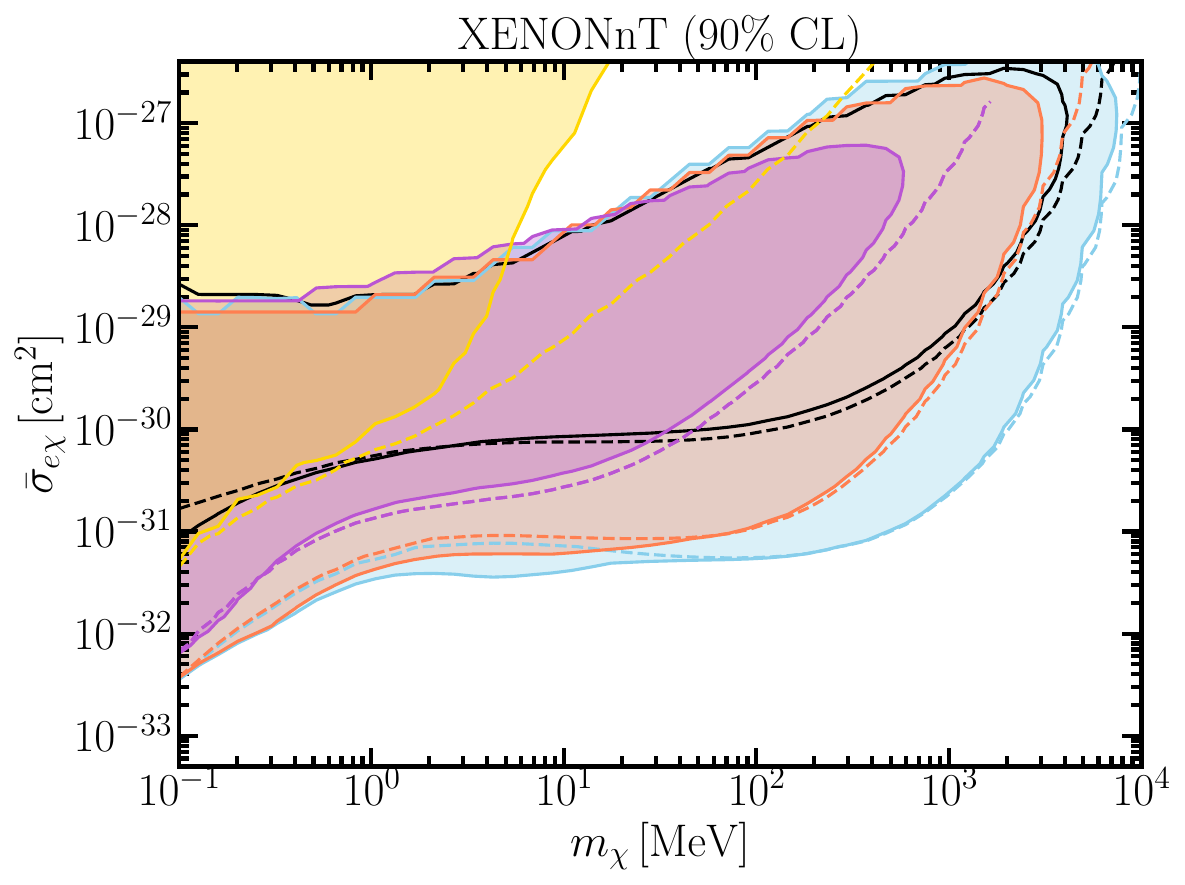}
    \includegraphics[width=0.49\linewidth]{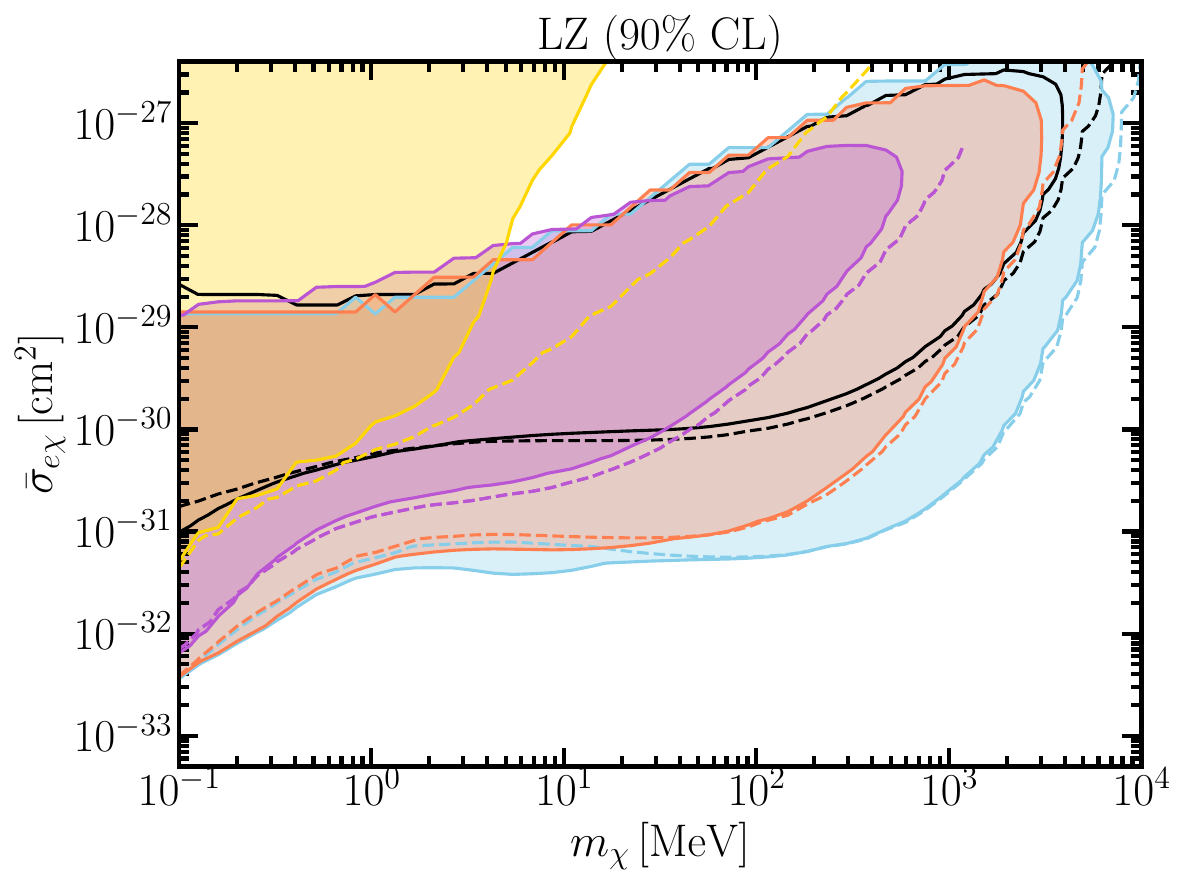}
    \includegraphics[width=0.49\linewidth]{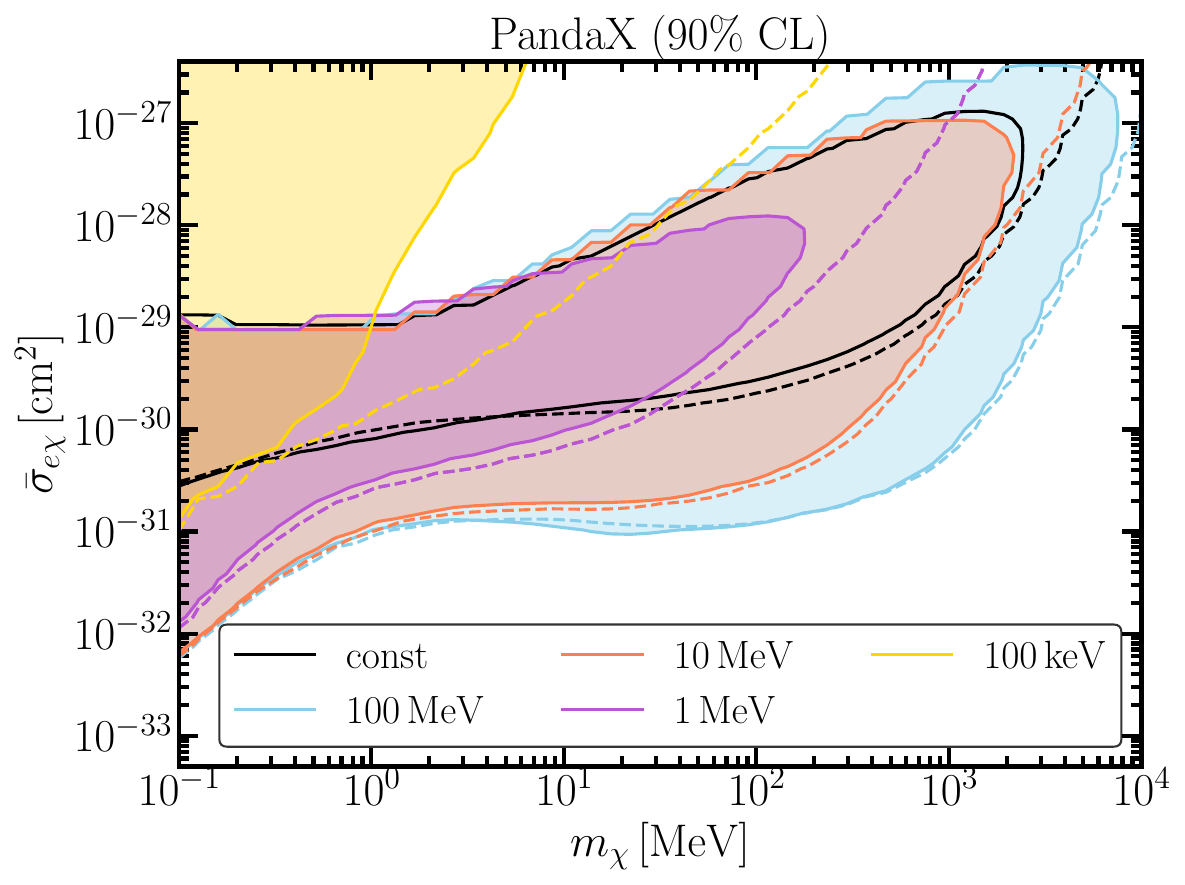}
    \caption{
    Constraints (90\% CL) on parameter space for the vector mediator model with four values of the mediator mass $m_{Z'}=0.1$ (yellow), 1 (purple), 10 (orange), and $100\MeV$ (blue). The constraint assuming a constant cross-section (energy-independent) is shown in black. The dashed lines show the corresponding constraints in the absence of attenuation.
    }
    \label{fig:vecconstraints}
\end{figure}

\begin{figure}[!t]
    \centering
    \includegraphics[width=0.49\linewidth]{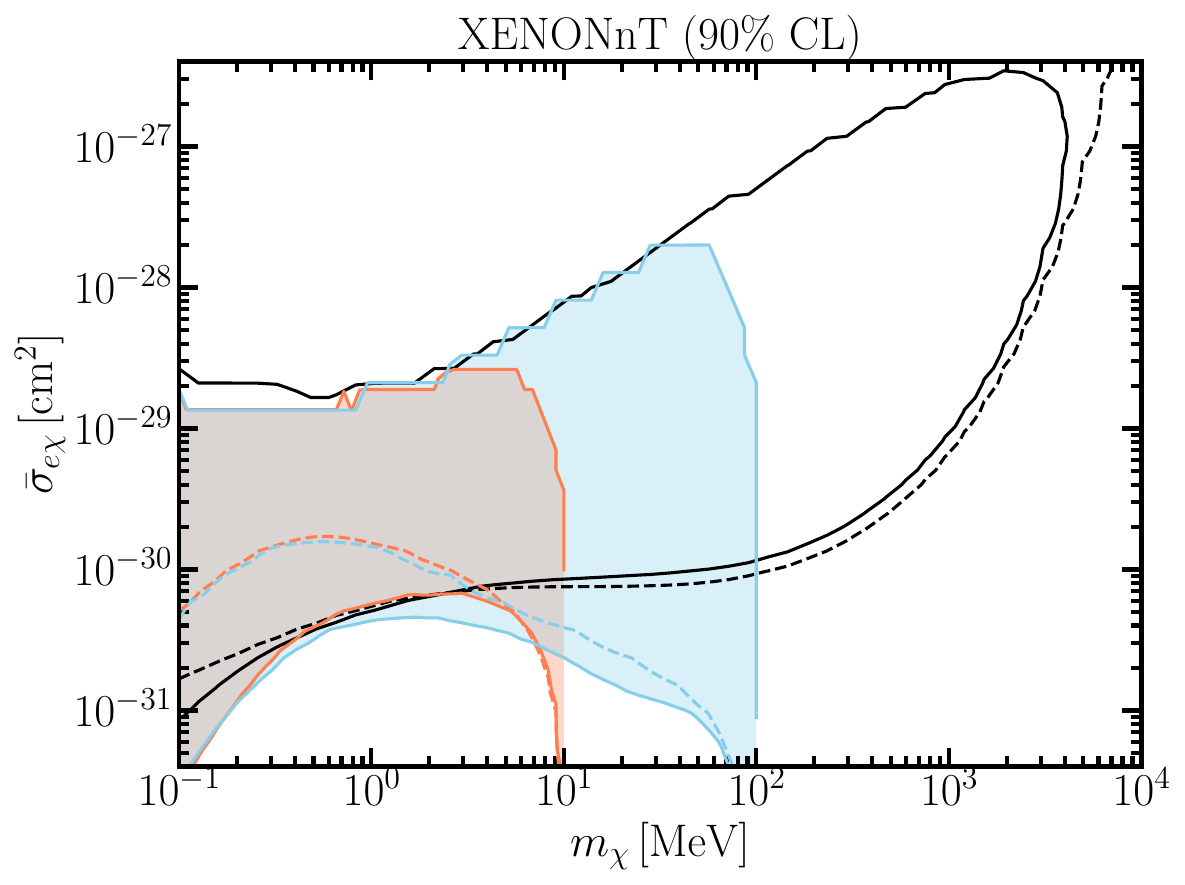}
    \includegraphics[width=0.49\linewidth]{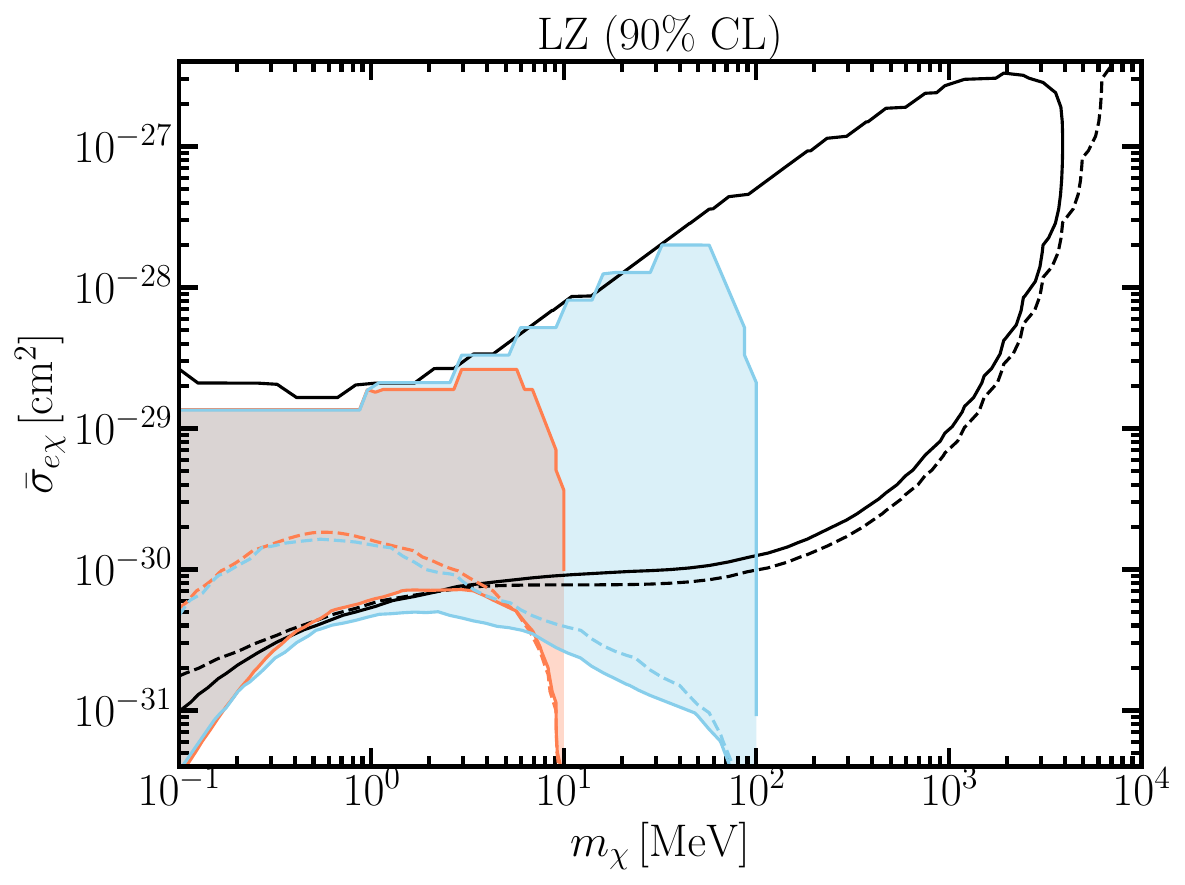}
    \includegraphics[width=0.49\linewidth]{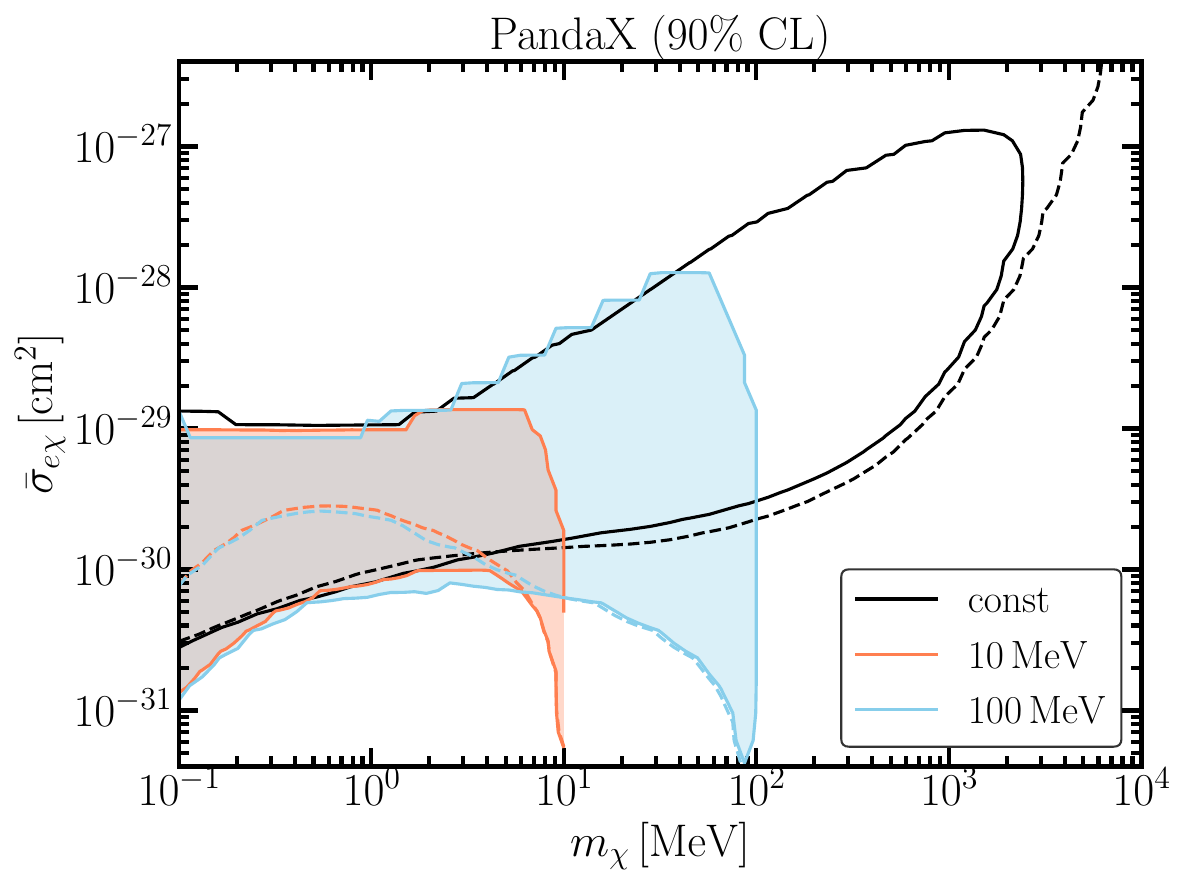}
    \caption{
    Constraints (90\% CL) on the parameter space for the scalar mediator model with two values of the mediator mass $m_\Phi=10$ (red) and $100\MeV$ (blue). The constraint assuming a constant cross-section (energy-independent) is shown in black. The dashed lines show the corresponding constraints in the absence of attenuation.
    }
    \label{fig:scaconstraints}
\end{figure}

We present the resulting exclusion regions for the three experiments for the vector and scalar mediators in Fig.\,\ref{fig:vecconstraints} and Fig.\,\ref{fig:scaconstraints} respectively. For reference, we also show constraints assuming a constant cross-section. The differences in these constraints compared to previous studies arise from relaxing the assumptions needed for the approximate analytical solution \cite{DeRomeri:2023ytt}.
The effect of mediator mass dependence on the constraints is found to be pronounced even without attenuation. In the case of a vector mediator, the full energy-dependent cross-section (coloured lines) tends to put stronger constraints than the energy-independent cross-section (black lines). This can be traced back to the enhanced cross-section for both the DM upscattering and interaction in the detector. We find that, in the absence of attenuation,  for heavier mediators, the exclusion region is larger, whereas for lighter mediators it shrinks considerably to lower values of $m_\chi$. The effect of attenuation is two-fold. Firstly, the exclusion region includes an upper limit (ceiling) on the interaction strength due to the complete attenuation of BDM. We note that some of the presented constraints exhibit numerical artifacts at the predicted ceiling edges, the precise position of which is highly sensitive to the DM coupling. These features appear due to finite parameter grid resolution. Secondly, we observe an enhancement of the constraints for lower values of $m_\chi$. This can be attributed to the down scattering of higher energetic BDM to lower energies more favourable for detection in the experiments.

For scalar-mediated interactions, the shape of the constraints is qualitatively distinct from the vector-mediated interactions. For most of the available parameter space, they tend to be weaker when compared to the constant cross-section case if attenuation is not considered. As the DM mass approaches the mediator mass, constraints become particularly strong. This can be traced to the resonant behaviour of the electron-DM cross-section, cf Eq.\,(\ref{eq:approx_echi_scalar}), and it further highlights the importance of the underlying Lorentz structure. On the other hand, the effects of attenuation are similar to the vector case with the typical upper limit and enhancement on the lower limit of the constraints. The latter is very pronounced for the scalar mediator case. We do not discuss limits for $m_\chi > m_\Phi$, which would introduce an annihilation channel of DM in the considered model.

As expected, the constraints cast by XENONnT and LZ are similar, as they employ a similar technology and are located at an approximately similar depth of $h_d \simeq 1.4\,\mathrm{km}$ from the surface. The results from PandaX tend to be slightly weaker compared to LZ and XENONnT. Since the former is located at a greater depth of $h_d \simeq 2.4\,\mathrm{km}$, the attenuation of the BDM flux is stronger. This leads to a weakening of both limits, the attenuation-induced upper limit, as well as the constraints on the lower mass of DM.  This is a result of the additional overburden, hence the overall loss of energy of BDM flux will not be locally compensated by overproducing DM with favourable kinetic energies for direct DM detection.
\begin{figure}[!]
    \centering
    \includegraphics[width=0.6\linewidth]{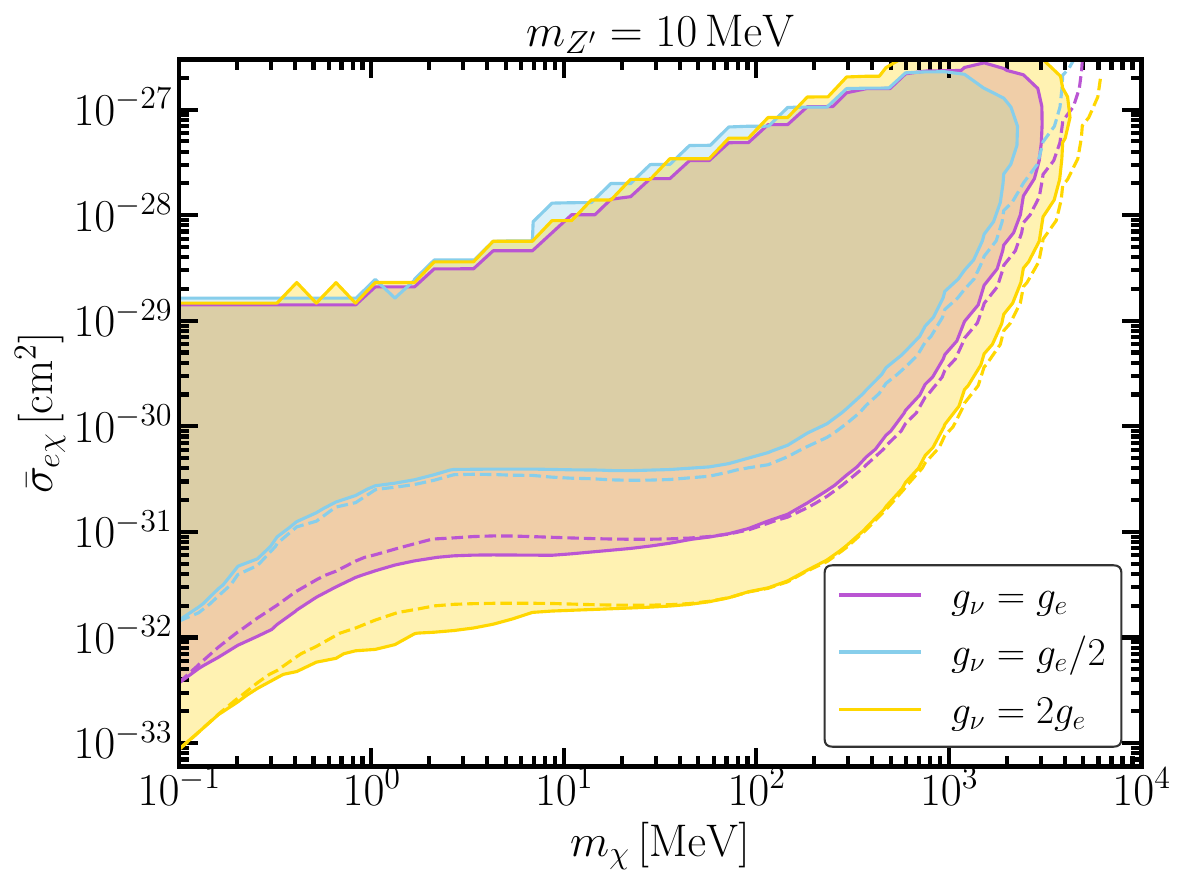}
    \caption{The effect of changing neutrino-DM coupling $g_\nu$ relative to $g_e$ on the constraints set by XENONnT. The attenuation ceiling is barely influenced, since the dominant effect of changing $g_\nu$ is changing the overall flux of BDM. The position of the ceiling is mildly sensitive to the overall flux but exponentially sensitive to $g_e$. The lower limits are dominated by the BDM flux and hence is affected by the change in $g_\nu$.}
    \label{fig:gnu}
\end{figure}

Let us emphasize again that the limits derived in Fig.\,\ref{fig:vecconstraints} and Fig.\,\ref{fig:scaconstraints} assume $g_\nu = g_e$. From a phenomenological point of view, these two couplings can be different and can yield different results for the same scenario. To demonstrate our point, we show in Fig.\,\ref{fig:gnu} the constraints for the specific case of vector mediator of mass $m_Z'=10\,\MeV$ for three values of $g_\nu=g_e,\,g_e/2,\, 2g_e$ in XENONnT. We find that the lower bound on the exclusion region shifts depending on the value of $g_\nu$, while the upper limit remains virtually unchanged. This is expected since the upper limit is fixed by the attenuation and depends on the value of $g_e$, whereas changing $g_\nu$ affects the BDM flux. 

Finally, we note that the limits we derived in this study are at relatively large DM cross-section values. This is a generic feature of boosted DM models as it requires DM to scatter twice with electrons and neutrinos. These cross-sections are excluded by other astrophysical observations and laboratory experiments\,\cite{Nguyen:2021cnb,Buen-Abad:2021mvc,SENSEI:2023zdf,DAMIC-M:2023gxo}. We do not display these additional constraints in our figures for clarity. However, a variety of additional constraints in these regions of parameter space require separate analysis since the exact nature of the constraint depends on the energy dependence of the DM-electron scattering cross-section, which has a significant impact on the resulting limits.

%%%%%%%%%
%%%%%%%%%%%%
\section{Summary and Conclusions}
\label{sec:conclusions}
%%%%%%%%%%%%%
%%%%%%%%%%%
Abundant neutrinos from past SN explosions in the Universe provide a testing ground for DM interactions with leptons. In this study, we analysed boosted DM due to upscattering with energetic neutrinos from the DSNB. Focusing on minimal leptophilic DM models, where DM interacts with neutrinos and electrons through a scalar or a vector boson,
we explored attenuation of the DM flux due to in-medium propagation as well as experimental signatures. Previous studies have focused on the simplifying assumption of constant DM interaction cross-sections and an analytic treatment of attenuation. We expand on the treatment of attenuation and discard the analytic treatment of attenuation in favour of a full numerical solution. We find this to be relevant even in the case of constant cross-section. We demonstrated that the inclusion of an energy-dependent cross-section significantly affects the boosted DM flux and the detection prospects.

We analyzed the constraints on boosted DM using recent data from XENONnT, LZ, and PandaX-4T large underground direct DM detection experiments. As expected, the rate of attenuation also depends on the details of the underlying DM model considered, differing considerably from the constant cross-section case.  With the effects of attenuation taken into account, we derived constraints on the model parameter space from these direct detection experiments and set new limits on DM neutrino and electron interactions for DM masses in the range $\sim (0.1, 10^4)$~MeV. By considering the energy dependence of DM interaction cross-sections, we demonstrated that resulting constraints can differ by multiple orders of magnitude compared to those found assuming constant cross-sections. These results showcase the inadequacies of the approximation of a constant cross-section as considered in previous studies.

Our work highlights the significance of the DSNB as an excellent target for probing neutrino-DM interactions. With the doping of the SK experiment with Gd, the expected sensitivity holds promise for DSNB detection in the near future. This discovery would open new avenues in neutrino astronomy. In light of this, our work is timely and sets the ground for further exploration of connections between this omnipresent astrophysical neutrino background and DM.

\acknowledgments
We thank John Beacom, Shunsaku Horiuchi, Anna Suliga for discussion.
We acknowledge the extensive use of \textsc{FeynCalc, NumPy, SciPy, Matplotlib} in this work\,\cite{Shtabovenko:2023idz,harris2020array,2020SciPy-NMeth,Hunter:2007}.
AD gratefully acknowledges the hospitality at APCTP during the program ``Dark Matter as a Portal to New Physics". AD was supported by Grant Korea NRF-2019R1C1C1010050.
TH acknowledges support from the IMPRS-PTFS. MS acknowledges the hospitality of the Network for Neutrinos, Nuclear Astrophysics, and Symmetries (N3AS) Physics Frontier Center of UC Berkeley where part of this work was carried out. TH and MS would like to thank Paul Frederik Depta for insightful discussions.
The work of VT was supported by the World Premier International Research Center Initiative (WPI), MEXT, Japan. VT acknowledges support the JSPS KAKENHI Grant 23K13109.

\bibliographystyle{JHEP}
\bibliography{bdm.bib}

\end{document}